\documentclass[prb,preprint]{revtex4-1}

\usepackage{amsmath}  
\usepackage{amsfonts} 
\usepackage{hyperref}
\hypersetup{colorlinks,urlcolor=blue,linkcolor=blue,citecolor=blue}
\usepackage{graphicx} 
\usepackage{subfigure}
\newcommand{\ud}{\mathop{}\negthinspace\mathrm{d}}
\newcommand{\uD}{\mathrm{D}}
\begin{document}
\title{Nonlinear coupling in an asymmetric pendulum}
\author{Qiuhan Jia}
\affiliation{School of Physics, Nanjing University, Nanjing, P.R. China, 210093}

\author{Yao Luo}
\altaffiliation[Permanent address: ]{Division of Engineering and Applied Science, California Institute of Technology, California, US, 91125}
\affiliation{School of Physics, Nanjing University, Nanjing, P.R. China, 210093}

\author{Huijun Zhou}
\affiliation{School of Physics, Nanjing University, Nanjing, P.R. China, 210093}

\author{Yinlong Wang}
\affiliation{School of Physics, Nanjing University, Nanjing, P.R. China, 210093}

\author{Jianguo Wan}
\affiliation{School of Physics, Nanjing University, Nanjing, P.R. China, 210093}

\author{Sihui Wang}\email{wangsihui@nju.edu.cn}
\affiliation{School of Physics, Nanjing University, Nanjing, P.R. China, 210093}

\date{\today}

\begin{abstract}
We investigate the nonlinear effect of a pendulum with the upper end fixed to an elastic rod which is only allowed to vibrate horizontally. The pendulum will start rotating and trace a delicate stationary pattern when released without initial angular momentum. We explain it as amplitude modulation due to nonlinear coupling between the two degrees of freedom. Though the phenomenon of conversion between radial and azimuthal oscillations is common for asymmetric pendulums, nonlinear coupling between the two oscillations is usually overlooked. In this paper, we build a theoretical model and obtain the pendulum's equations of motion. The pendulum's motion patterns are solved numerically and analytically using the method of multiple scales. In the analytical solution, the modulation period not only depends on the dynamical parameters, but also on the pendulum's initial releasing positions, which is a typical nonlinear behavior. The analytical approximate solutions are supported by numerical results. This work provides a good demonstration as well as a research project of nonlinear dynamics on different levels from high school to undergraduate students.
\end{abstract}

\maketitle

\section{Introduction}
The ideal trajectory of a two-dimensional asymmetric pendulum is a Lissajous figure, the superposition of two independent simple harmonic motion (SHM). A common example of asymmetric pendulum is a ``Y-suspended" pendulum which was invented twice for scientific and recreational purposes.\cite{Whitaker2004,Greenslade2003,Whitaker1991} When the frequency ratio of the two oscillations $\omega_1/\omega_2$  is a rational number, the trajectory will be stationary. Otherwise, when $\omega_1/\omega_2$ is not exactly a rational number, the motion is quasi-periodic and the trajectory varies with time due to a growing phase drift.\cite{Jorge2011} Singh et al. described that in an asymmetric two-dimensional pendulum, the quasi-periodic motion shifts from planar to elliptical and back to planar again.\cite{Singh2018} The period that the pendulum returns to planar motion was related to the strength of symmetry breaking introduced with an additional spring.

Once linear coupling is introduced to an oscillation system with two degrees of freedom, for instance, by connecting two identical pendulums with a weak spring, two normal modes are formed, whose frequencies are usually different from the pendulums' natural frequencies.\cite{Feynman} The resultant motion described by linear superposition of the two normal modes with slightly different frequencies may give rise to a ``beat" motion, in which the amplitudes of the pendulums varies slowly and energy is transferred cyclically between the two pendulums.   

 If we look at a real asymmetric pendulum, perfect independent motions are quite unlikely to happen. The phenomenon of conversion between radial and azimuthal oscillations is common for asymmetric pendulums. However, nonlinear coupling between the two oscillations is usually overlooked. In this paper, we study the dynamics of a quasi-two-dimensional pendulum with weak nonlinear coupling. 
\begin{figure}[t]
    \begin{center}
	\subfigure[]{
    \includegraphics[height=0.4\textwidth]{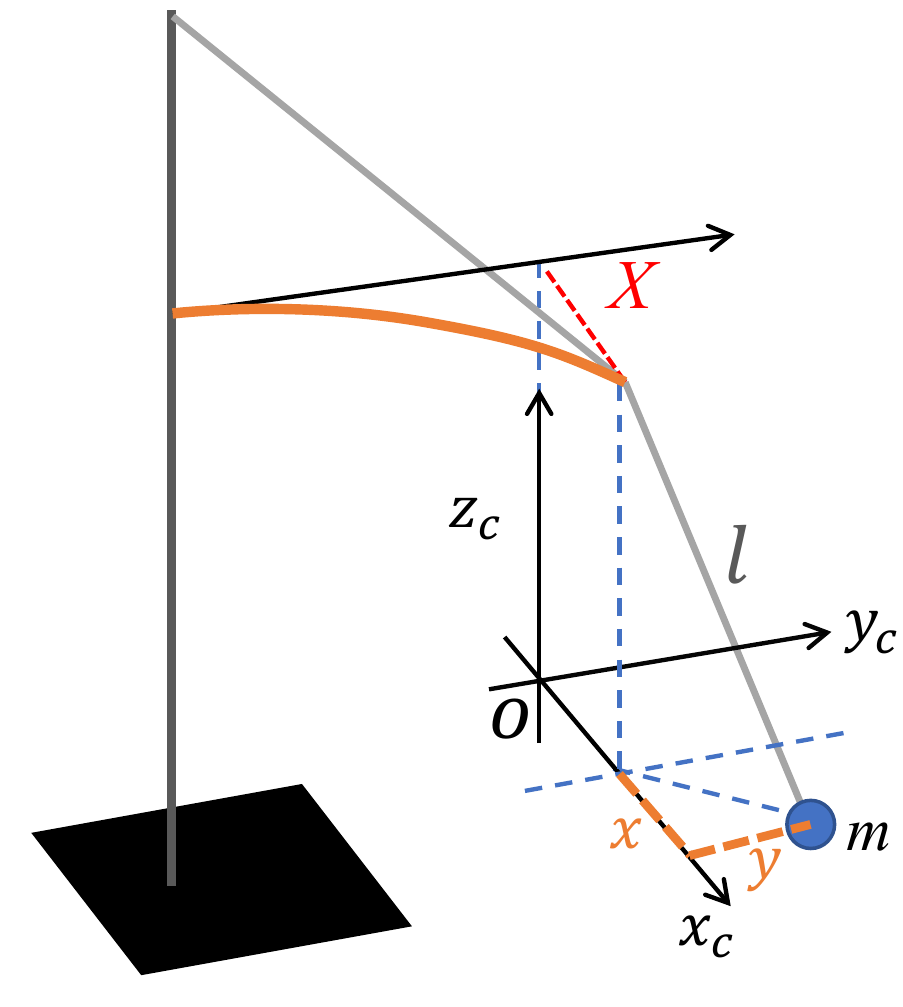}
	\label{fig:coordinates}
    }
    \subfigure[]{
    \includegraphics[height=0.35\textwidth]{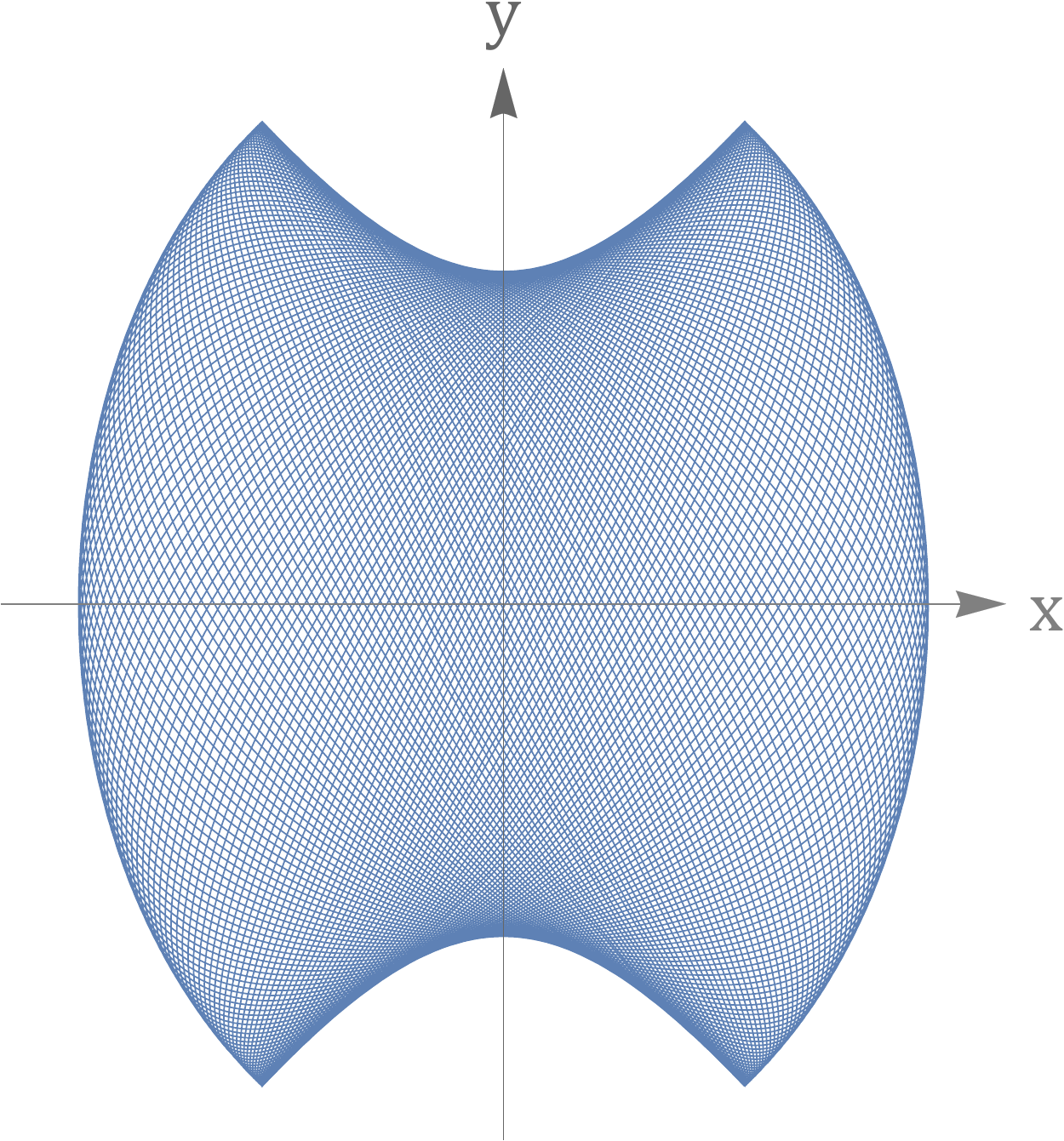}
	\label{fig:example}
    }
    \subfigure[]{
    \includegraphics[height=0.35\textwidth]{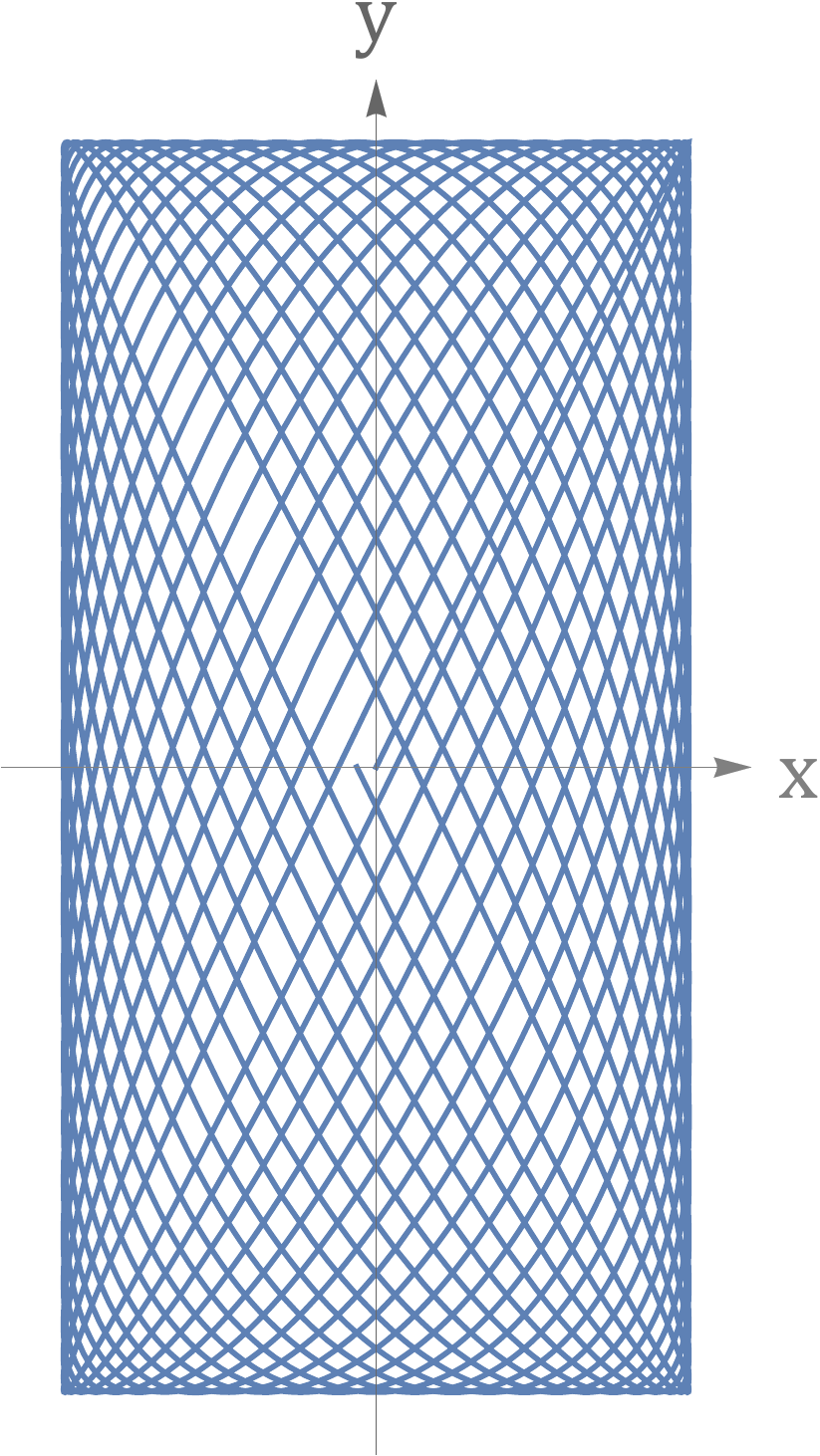}
	\label{fig:lissajous}
	}
    \caption{(a)The asymmetric pendulum. The upper end of the string is fixed to an elastic rod only allowed to vibrate horizontally. The generalized coordinates are also defined.  (b) A typical pattern derived from numerical results. (c) A Lissajous figure for comparison.}        
    \end{center}
\end{figure}

The pendulum is shown in Fig.~\ref{fig:coordinates}. Suspend a bob on a string from the end of an elastic rod. The other end of the rod is supported with another taut string to avoid vertical deflection. If the pendulum is released in a plane parallel to the rod, the radial oscillation will spontaneously convert into a motion that shifts from planar to elliptical motion and back to planar again. An experimental video clip and an animation are included in supplementary material 1.\cite{supplementary_material_1} A typical pattern of this quasi-2D pendulum's motion is shown in Fig.~\ref{fig:example}. At first sight, the pattern resembles a proceeding Lissajous figure with varying phase. A Lissajous figure is plotted in Fig.~\ref{fig:lissajous} for comparison. We see that in a Lissajous figure, the amplitudes in $x$ and $y$ directions are fixed, while the quasi-2D pendulum amplitudes in both directions change alternately.  Amplitude change and energy transfer between the azimuthal ($x$) and radial ($y$) directions are apparently consequences of coupling between the two degrees of freedom.

In this paper, we build a theoretical model and simplify the pendulum’s equations of motion into two-dimensional. The pendulum’s motion is solved numerically and analytically using the method of multiple scales. The difficulty in this problem is that the coupling is nonlinear, and the motion cannot be simply decomposed into two normal modes. Fortunately, nonlinearity in this problem is small and can be treated as perturbation to the two independent oscillations on radial and azimuthal directions. In the analytical solution, each motion is comprised of two oscillations with slightly different frequencies. We introduce the nonlinear modulation period and modulation depth to describe the feature of the motion. The modulation period and modulation depth we derived not only depends on dynamical parameters, but also on the pendulum’s initial releasing positions. The analytical solutions are consistent with numerical solutions with good accuracy when nonlinear effect is weak.

This problem has aroused extensive interest among students as a popular competition problem in the 2018 International Young Physicists’ Tournament and  China Undergraduate Physics Tournament.\cite{IYPT}
The phenomenon described in this problem is common in many two dimensional asymmetric pendulums. The advantage of this experimental apparatus is that it has appealing visual effects and the strength of coupling and other oscillation parameters are  controllable thus can easily be compared to theoretical results. The solution in this paper provides a theoretical explanation for the phenomenon. This work provides a good demonstration experiment as well as a research project from high school to undergraduate students. 

\section{Theoretical Model}
\subsection{Equation of Motion}
We write the Lagrangian of the pendulum in terms of Cartesian coordinates. As shown in Fig.~\ref{fig:coordinates}, the origin $O$ is taken as the bob’s equilibrium position when the rod has no deflection. The bob’s coordinates $x_c, y_c, z_c$ are taken with respect to $O$. $x$ and $y$ are the bob’s relative coordinates with respect to the rod’s oscillatory end. For small deflection, the horizontal deflection of the end is approximately one dimensional and denoted as $X$. Therefore,
\begin{align}
	x_c&=X+x,\\
	y_c&=y,\\
		z_c&=l-\sqrt{l^2-(x^2+y^2)}\simeq\frac1{2l}(x^2+y^2),\label{eq:coordinate_zc}
\end{align}
where $l$ is the string length. In this problem, we consider a hard rod whose natural frequency is much higher than that of the pendulum.  Moreover, as the rod is slender and the deflection is small, we can model the rod using Euler-Bernoulli beam theory. More specifically, the rod is treated as a cantilever beam bent by a force at the free end.\cite{carrera2011} Actually, the deflection at any point is approximately proportional to the force, and the strain energy and kinetic energy are both quadratic. Hence, The total kinetic energy of the bob and the rod is
\begin{align}
	T&=\frac{1}{2}m(\dot{x_c}^2+\dot{y_c}^2+\dot{z_c}^2)+\frac{1}{2}M^*\dot{X}^2,
\end{align}
where $m$ is the mass of the bob, $\frac{1}{2}M^*\dot{X}^2$ is the rod's kinetic energy expressed in term of an effective mass $M^*$.
 
The potential energy is
\begin{align}
	V&=mg z_c+\frac{1}{2}k X^2,
\end{align}
where the rod's potential energy is given in term of an effective elastic coefficient $k$.

The Lagrangian is therefore
\begin{align}
	\label{eq:Lxy}
	L=\frac{1}{2} m((\dot{x}+\dot{X})^2+\dot{y}^2)+\frac{1}{2} M^*\dot{X}^2-\frac{1}{2} k X^2-\frac{mg}{2l}\left(x^2+y^2\right)+\frac{m}{2l^2}\left( x\dot{x}+y\dot{y}\right)^2.
\end{align}

The equations of motion obtained using Euler-Lagrange (E-L) equations with $x$, $y$ and $X$ are
\begin{subequations}
	\label{eq:3D equations}
	\begin{align}
	\ddot{X}+\ddot{x}+\omega_y^2x+{\color{red}\frac1{l^2}(x\dot{x}^2+x\dot{y}^2+x y\ddot{y}+x^2\ddot{x})}
	=0,\label{eq:3D equations x}\\
	\ddot{y}+\omega_y^2y+{\color{red}\frac1{l^2}(y\dot{x}^2+y\dot{y}^2+x y\ddot{x}+y^2\ddot{y})}
	=0,\label{eq:3D equations y}\\
	(1+\gamma)\ddot{X}+\ddot{x}+\omega_X^2X=0,
	\end{align}
\end{subequations}
where $\omega_y^2=\frac{g}{l},\,
\omega_X^2=\frac{k}{m},\,\gamma=\frac{M^*}{m}$.

Rearrange the highlighted terms, Eqs.~(\ref{eq:3D equations x}) and (\ref{eq:3D equations y}) can be rewritten as  
\begin{subequations}
	\begin{align}
	\ddot{X}+\ddot{x}+\omega_y^2x+{\color{red}\frac{\ud{}}{\ud{}t}\frac{\partial \frac12{v_z}^2}{\partial\dot{x}}-\frac{\partial \frac12{v_z}^2}{\partial x}}
	=0,\\
	\ddot{y}+\omega_y^2y+{\color{red}\frac{\ud{}}{\ud{}t}\frac{\partial \frac12{v_z}^2}{\partial\dot{y}}-\frac{\partial \frac12{v_z}^2}{\partial y}}
	=0.
	\end{align}
\end{subequations}

We see that all the nonlinear terms come from the vertical motion which is usually neglected concerning the pendulum's short-term behavior. When long-term behavior is considered, the effects of these small terms accumulate over time, and they modulate the amplitudes on $x$ and $y$ directions periodically. We will treat the effects of these small terms as perturbation. Before doing so, we firstly simplify the pendulum's motion into two dimensional by considering the ``undisturbed" solution.

\subsection{Simplification: Two Dimensional Model}
When the nonlinear terms are ignored  in Eq.~(\ref{eq:3D equations}), the pendulum's equations of motion for the undisturbed system (the generating system) is reduced to 
\begin{subequations}
	\label{eq:generating equations}
	\begin{align}
	(\ddot{X}+\ddot{x})+\omega_y^2x&=0,\label{eq:x1}\\
	\ddot{y}+\omega_y^2y&=0,\label{eq:y}\\
	(1+\gamma)\ddot{X}+\ddot{x}+\omega_X^2X&=0.\label{eq:X}
	\end{align}
\end{subequations}

In Eq.~(\ref{eq:y}), we see that $y$ is independent of $x$ and the its solution is simple harmonic motion. $x$ and $X$ are still coupled. 
Write the trial solution as 
\begin{align}
	x&=x_0e^{\lambda t},&X&=X_0e^{\lambda t},\label{eq:trial}
	\end{align}
and substitute them into Eqs.~(\ref{eq:generating equations}), we find that
	\begin{subequations}
	\begin{align}
	\lambda^2X_0+(\lambda^2+\omega_y^2)x_0&=0,\\
	\left((1+\gamma)\lambda^2+\omega_X^2\right)X_0+\lambda^2x_0&=0.
	\end{align}
\end{subequations}
Nontrivial solution exists only if the determinant equals zero. Therefore, 
\begin{subequations}
	\begin{align}
	\left|
	\begin{array}{cc}
		\lambda^2 & \lambda^2+\omega_y^2\\
		(1+\gamma)\lambda^2+\omega_X^2 & \lambda^2
		\end{array}
	\right|=0,
	\end{align}
that is,
	\begin{align}
	\gamma\lambda^4+(\omega_X^2+(1+\gamma)\omega_y^2)\lambda^2+\omega_X^2\omega_y^2=0.
	\end{align}
\end{subequations}
The solutions of this equation are

\begin{subequations}
	\label{eq:two omegas}
	\begin{align}
	\lambda_-^2 &\simeq -\left(1-\kappa\right)\omega_y^2 \equiv -\omega_x^2,\\
	\lambda_+^2 &\simeq -\frac{1}{\gamma}\left(1+\kappa\right)\omega_X^2 \equiv -{\omega'}^2,
	\end{align}
\end{subequations}
where $\kappa\equiv\frac{mg}{kl}\ll 1$ and we have made a first-order approximation. $\omega_x$ is an effective natural frequency of the bob's azimuthal oscillation which is slightly lower than the pendulum's natural frequency and $\omega'$ is a much higher frequency of the magnitude order of the rod's natural frequency.

Substitute each $\lambda$ in Eqs.~(\ref{eq:two omegas}) into Eq.~(\ref{eq:x1}), we have
\begin{align}
	\frac{X_0}{x_0}&=\frac{\omega_y^2+\lambda^2}{-\lambda^2},\notag\\
	&=\begin{cases}
	\frac{\kappa}{1-\kappa}&\text{, when taking $\lambda_-.$}\\
	\frac{\kappa}{1+\kappa}\gamma-1&\text{, when taking $\lambda_+.$}
	\end{cases}\label{eq:lambda+- cases}
\end{align}
For the solution with lower frequency $\omega_x$, the bob's oscillation amplitude is much greater than that of the rod's, since $\kappa\ll 1$. For the solution with high frequency $\omega'$, they are of the same magnitude. Normally the bob's displacement is much greater than the displacement of the rod's end, therefore we will neglect the high frequency solution. Thus, Eq.~(\ref{eq:lambda+- cases}) becomes
\begin{align}
X_0=\frac{\kappa}{1-\kappa} x_0.
\end{align}
So,
\begin{align}
X=\frac{\kappa}{1-\kappa} x.\label{eq:equivalent X}
\end{align}
Substitute Eq.~(\ref{eq:equivalent X}) into Eqs.~(\ref{eq:generating equations}), the equations are reduced to two dimensional as
\begin{subequations}
	\begin{align}
	\ddot{x}+\omega_x^2x&=0,\\
	\ddot{y}+\omega_y^2y&=0.
	\end{align}
\end{subequations}
Apparently, for the "undisturbed system", the asymmetry is simply described by the difference in frequencies of azimuthal and radial oscillations introduced by the elastic rod.

Retaining the nonlinear coupling terms, we obtain the following equations of motion 
\begin{subequations}
\label{eq:simplified equations}
	\begin{align}
	\ddot{x}+(1-\kappa)x&=-(1-\kappa)(x\dot{x}^2+x\dot{y}^2+x y\ddot{y}+x^2\ddot{x}),\label{eq:x final equation}\\
	\ddot{y}+y&=-(y\dot{y}^2+y\dot{x}^2+x y\ddot{x}+y^2\ddot{y}).\label{eq:simplified equations y}
	\end{align}
\end{subequations}
In this equation and below, for simplicity, the variables $x$ and $y$ represent $x/l$ and $y/l$, and $t$ represents $\omega_y t$.

Notice that in Eqs.~(\ref{eq:simplified equations}), there is only one parameter, the coupling coefficient $\kappa$, which determines the frequency difference between the two directions. Also note that $\kappa$ is independent of the effective mass of the rod $M^*$. The mass of the rod has little effect on this system because the kinetic energy of the rod is negligible compared to its elastic potential energy, when we only consider the low frequency motion.

\section{Numerical and Analytical Results}
\subsection{Numerical Result}
\begin{figure}[t]
	\begin{center}
	\subfigure[]{\label{fig:x_t}
	\includegraphics[width=0.8\textwidth,origin=c]{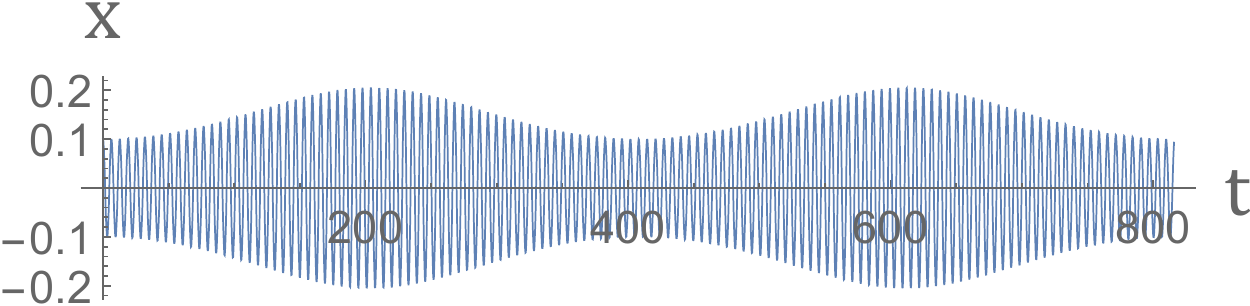}
	}
	\subfigure[]{\label{fig:y_t}
    \includegraphics[width=0.8\textwidth,origin=c]{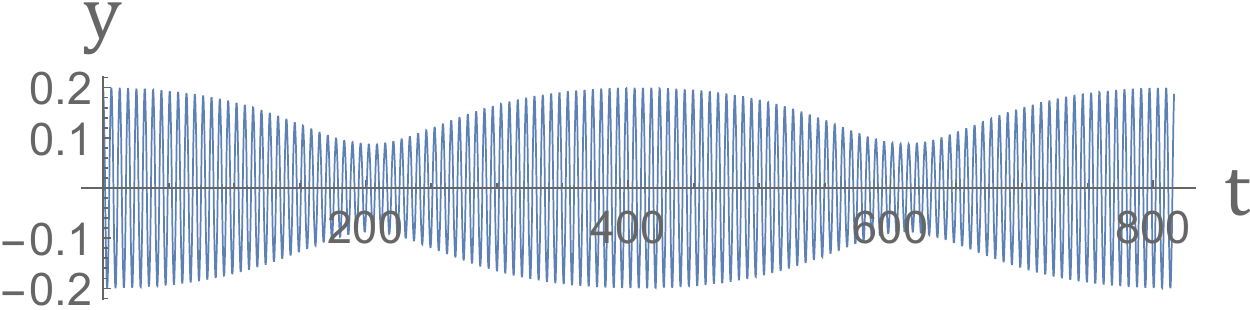}
    }
    
    \subfigure[]{\label{fig:xy}
	\includegraphics[width=0.35\textwidth,origin=c]{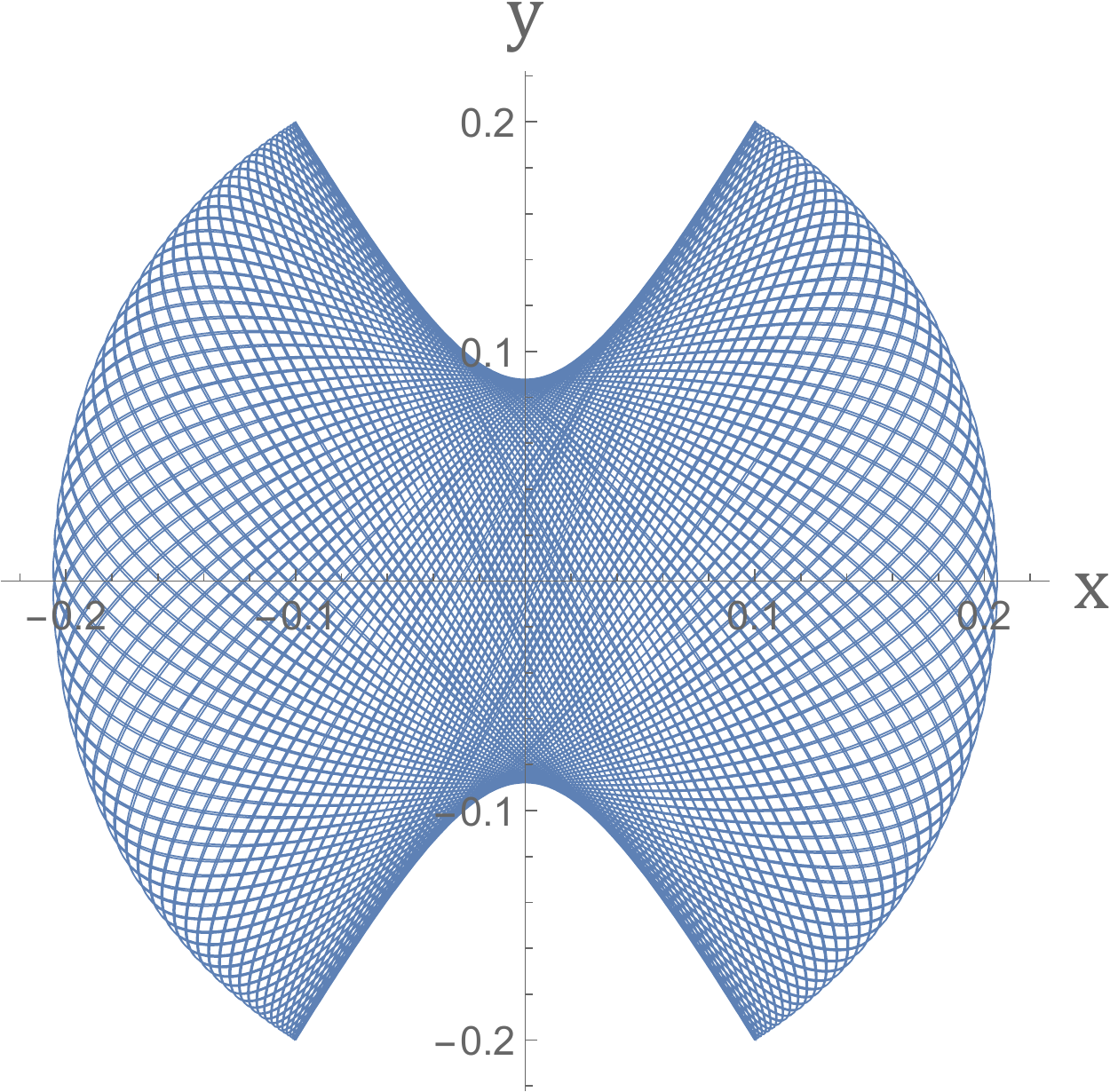}
	}
	\subfigure[]{\label{fig:frequency spectrum}
	\includegraphics[width=0.45\textwidth,origin=c]{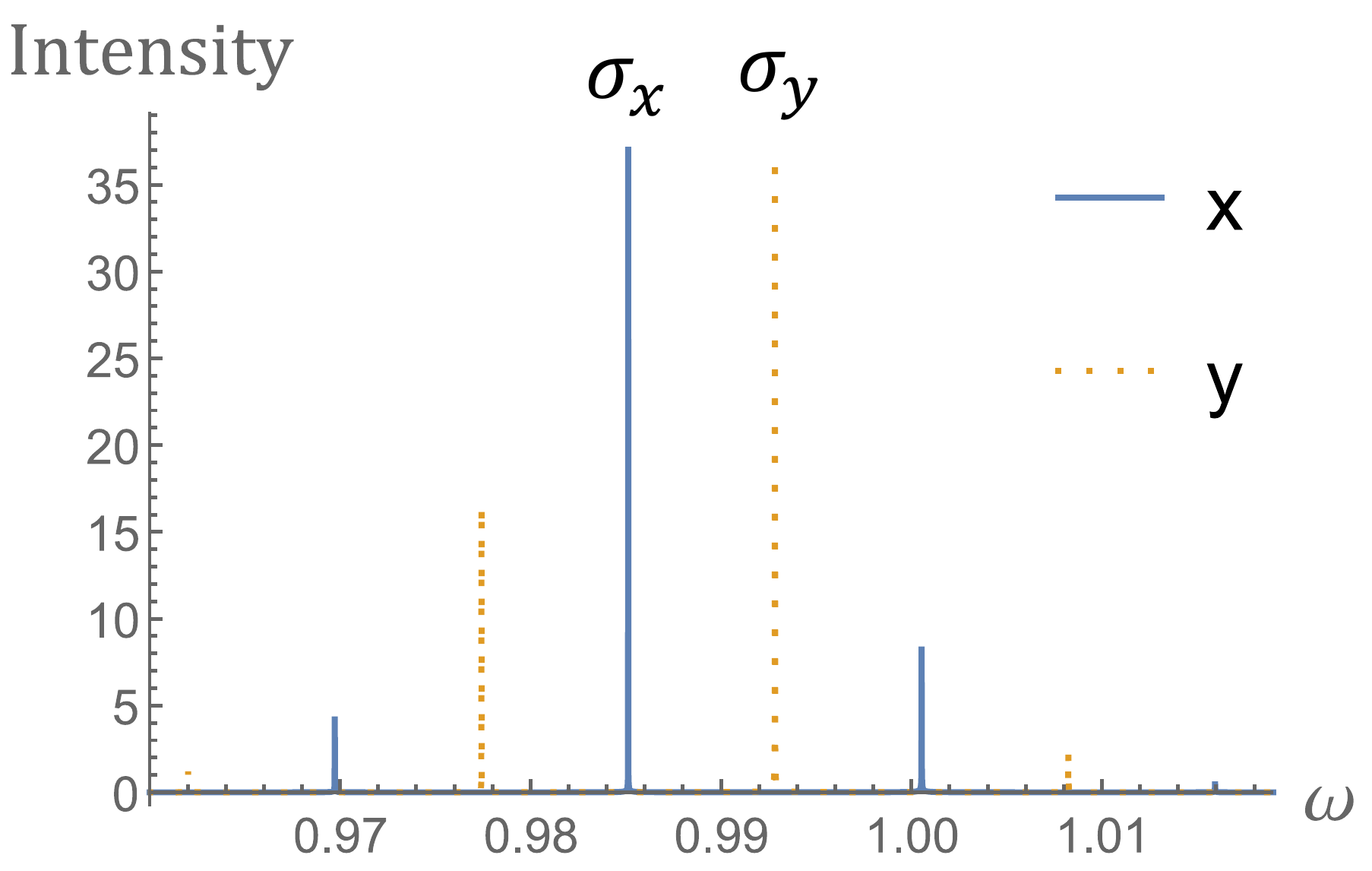}
	}
	\caption{(a) and (b) show the oscillations on $x$ (the azimuthal) and $y$ (the radial) directions. (c) shows the trajectory projected on the horizontal plane. (d) is the frequency spectrum of motions in $x$ and $y$ directions.}
	\label{fig:Simulative trajectory}
	\end{center}
\end{figure}

Applying NDSolve method\cite{NDSolve} in Wolfram Mathematica to solve Eqs.~(\ref{eq:simplified equations}), we obtain the results as shown in the Fig.~\ref{fig:Simulative trajectory}.
In the calculation, $\kappa=0.01$, and the pendulum is released from $x(0)=0.1,\,y(0)=0.2,\,x'(0)=y'(0)=0$. Figs.~\ref{fig:x_t} and \ref{fig:y_t} show the oscillations on $x$ (the azimuthal) and $y$ (the radial) directions respectively. In the figures, the oscillation on $x$ and $y$ directions resembles that of a ``beat'' motion, in which the amplitudes on $x$ and $y$ directions are modulated periodically. The time span in the figures is twice the amplitude modulation periods $T$. Within $2T$, the pendulum returns to initial planar motion and the trajectory forms a complete stationary pattern as shown in Fig.~\ref{fig:xy}. We will see that the pendulum's phase evolution period is $2T$.
\begin{figure}[t]
	\begin{center}
	\includegraphics[width=\textwidth]{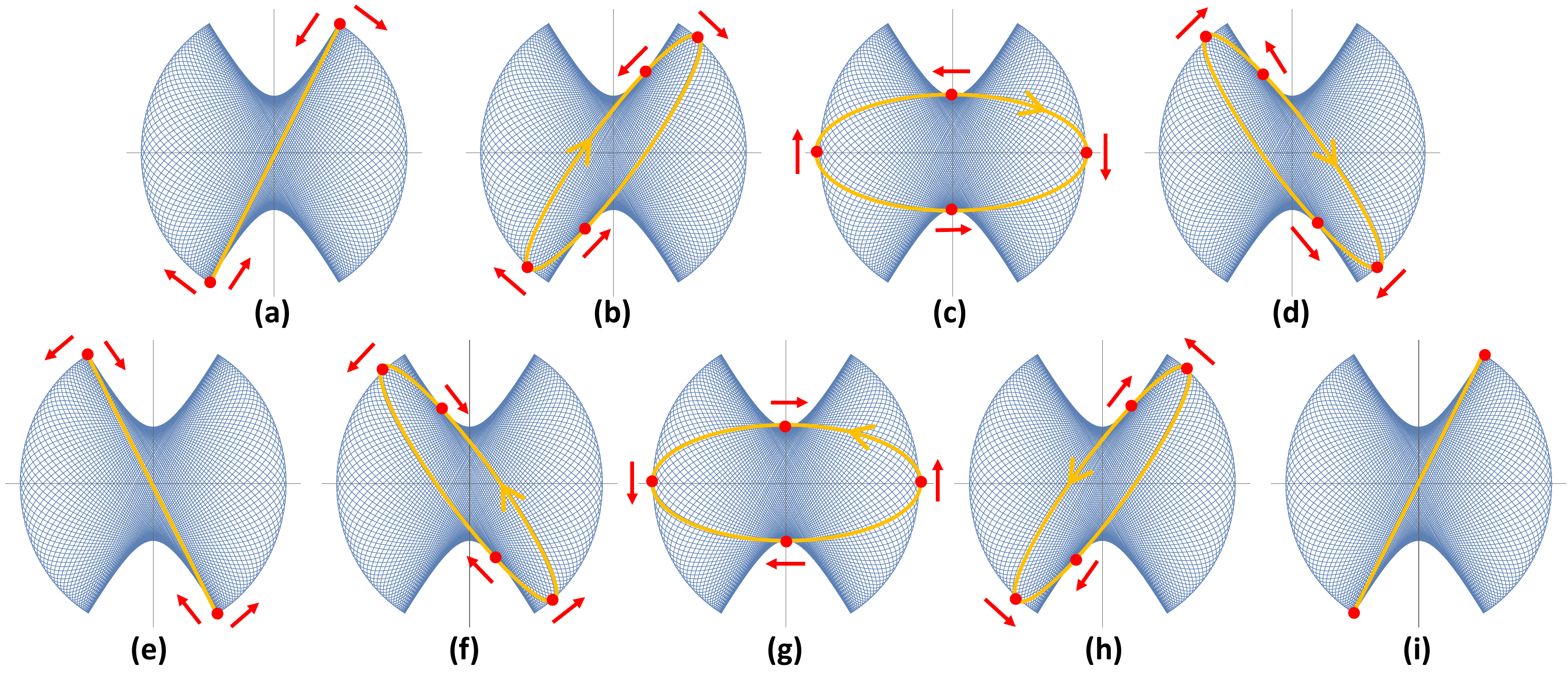}
	\caption{These figures show how the motion evolves in a period. The yellow ellipses depict short term motion and the arrows on the ellipses depict the rotation direction of the bob. The red points are the points of tangency and the red arrows beside the points show how these points move. In (a), the bob is released from the upper-right point. Then the bob rotates "elliptically". The direction of rotation is clockwise from (a) to (e) and counter-clockwise from (e) to (i), which is the same as the rotation of the ellipse axes.}
	\label{fig:evolution}
	\end{center}
\end{figure}

To illustrate the phase and amplitude evolution more clearly, we plot the evolution diagrams within $2T$ in Fig.~\ref{fig:evolution}. The background blue pattern is the same as in Fig.~\ref{fig:xy} which clearly shows the stationary envelope of the trajectory. And the yellow curves are the short term trajectory as the bob moves back and forth within one oscillation. Within each oscillation, the trajectory is nearly closed since the rate of procession is very small. So we can approximate them as ellipses gradually rotating and transforming to depict the short term motion. The red points are the points of tangency of the ellipse and the envelope curve, and the arrows show how the points of tangency move. After the bob is released at the upper-right point in Fig.~\ref{fig:evolution}(a), the planar motion gradually becomes “elliptical” and both the major axis and the bob rotate clockwise, see the animation.\cite{supplementary_material_1} The proceeding rate increases and reaches maximum when the major axis of the ellipse passes the x-axis in Fig.~\ref{fig:evolution}(c). Then the proceeding rate begins to decrease and finally the oscillation returns to planar in the position symmetric to the initial position at time $T$, as shown in Fig.~\ref{fig:evolution}(e). At the time, the amplitude for $y$ oscillation returns to maximum, and the amplitude for $x$ oscillation returns to minimum. Afterwards, the pendulum reverses its rotation counter-clockwise from Fig.~\ref{fig:evolution}(e) to Fig.~\ref{fig:evolution}(i). As we haven’t considered the effect of damping, the motion is quasi-periodic: the pendulum repeats this motion pattern cyclically similar to the moving Lissajous figure with slightly different frequencies. Hence, the duration from Fig.~\ref{fig:evolution}(a) to Fig.~\ref{fig:evolution}(i) is the period of phase modulation $2T$.

Then we perform Fourier transformation to $x$ and $y$ oscillations and obtain the frequency spectra in Fig.~\ref{fig:frequency spectrum}. In the figure, the spectra of $x$ and $y$ oscillations are indicated in solid blue line and dashed yellow line respectively. We see that both oscillation spectra comprise of several discrete peaks which are evenly spaced. We will show analytically that the spacing is twice the frequency difference of the main peaks $\sigma_y-\sigma_x$. Both the main peak frequencies of $x$ and $y$ oscillations are slightly lower than the natural frequency of undisturbed motion, which equals 1 in dimensionless Eq.~(\ref{eq:simplified equations y}). The discrete spectra of each oscillation with spacing  $2(\sigma_y-\sigma_x)$ give rise to amplitude modulation featured as the “beat” phenomenon with period $T=\frac{2\pi}{2(\sigma_y-\sigma_x)}$. The main peak frequency difference $\sigma_y-\sigma_x$ gives rise to phase shifting with period $2T=\frac{2\pi}{\sigma_y-\sigma_x}$.

Moreover, the results obtained from the simplified Eqs.~(\ref{eq:simplified equations}) show no observable difference when we substitute the same parameters into the original equations (\ref{eq:3D equations}). This proves that our simplification is reasonable. To see how nonlinear coupling affects the pendulum’s motion patterns, we utilize an analytical method for further study.

\subsection{The Method of Multiple Scales}
The method of multiple scales was first introduced by Peter A. Sturrock\cite{Sturrock} in 1957 and developed by Nayfeh\cite{Nayfeh1993, Nayfeh1995, Nayfeh2000} and others.\cite{Kevorkian1996, Bender1999} The underlying idea of the method of multiple scales is to regard the motion as a superposition of motions in multiple time scales which are independent variables.
Firstly, we introduce independent time scale variables according to
\begin{align}
T_n=\varepsilon^n t\quad(n=0,1,2,\cdots),
\end{align}
where $\varepsilon$ is a dimensionless small quantity. It follows that the derivatives with respect to $t$ become expansions in terms of the partial derivatives with respect to $T_n$ according to
\begin{align}
\label{eq:derivitive expansion}
\frac{\ud}{\ud t}&=\frac{\partial}{\partial T_0}\frac{\ud T_0}{\ud t}+\frac{\partial}{\partial T_1}\frac{\ud T_1}{\ud t}+\frac{\partial}{\partial T_2}\frac{\ud T_2}{\ud t}+\cdots,\notag\\
&=\frac{\partial}{\partial T_0}+\varepsilon\frac{\partial}{\partial T_1}+\varepsilon^2\frac{\partial}{\partial T_2}+\cdots,\notag\\
&=\uD_0+\varepsilon \uD_1+\varepsilon^2 \uD_2+\cdots,
\end{align}
where $\uD_n\equiv\frac{\partial}{\partial T_n}$. One assumes that the solution can be represented by an expansion in the form
\begin{align}
x(t,\varepsilon)=\sum_{n=1}^{m+1}\varepsilon^n x_n(T_0,T_1,T_2,\cdots,T_m),\label{eq:x expansion}
\end{align}
where m is the the order to which we need to carry out the expansion. Here we take $m=2$.

Substituting Eq.~(\ref{eq:derivitive expansion}) and Eq.~(\ref{eq:x expansion}) into Eq.~(\ref{eq:simplified equations}) and equating the coefficients of $\varepsilon$ to the power of 1, 2 and 3 separately, we obtain
\begin{subequations}\label{eq:1 order equation}
\begin{align}
\uD_0^2 x_1+(1-\kappa)x_1 &=0,\label{eq:1 order x equation}\\
\uD_0^2 y_1+y_1 &=0,\label{eq:1 order y equation}
\end{align}
\end{subequations}
\begin{subequations}\label{eq:2 order equation}
\begin{align}
\uD_0^2 x_2+(1-\kappa)x_2 &=-2\uD_1\uD_0x_1,\label{eq:2 order equation x}\\
\uD_0^2 y_2+y_2 &=-2\uD_1\uD_0y_1,\label{eq:2 order equation y}
\end{align}
\end{subequations}
and
\begin{subequations}\label{eq:3 order equation}
\begin{align}
\uD_0^2 x_3+(1-\kappa)x_3 =&-2\uD_2\uD_0x_1 -2\uD_1\uD_0x_2 -\uD_1^2x_1\notag\\
&-(1-\kappa) \big[x_1(\uD_0x_1)^2 +x_1(\uD_0y_1)^2 +x_1y_1\uD_0^2y_1 +x_1^2\uD_0^2x_1\big],\label{eq:3 order equation x}\\
\uD_0^2y_3+y_3 =&-2\uD_2\uD_0y_1-2\uD_1\uD_0y_2 -\uD_1^2y_1\notag\\
&-\big[y_1(\uD_0y_1)^2 +y_1(\uD_0x_1)^2 +y_1x_1\uD_0^2x_1 +y_1^2\uD_0^2y_1\big].\label{eq:3 order equation y}
\end{align}
\end{subequations}
The solution of Eqs.~(\ref{eq:1 order equation}) is
\begin{subequations}
\label{eq:1 order solution}
\begin{align}
x_1 &=A(T_1,T_2)e^{i(1-\frac12\kappa)T_0}+cc,\\
y_1 &=B(T_1,T_2)e^{i T_0}+cc,
\end{align}
\end{subequations}
where $A(T_1,T_2)$ and $B(T_1,T_2)$ are complex amplitudes to be solved and $cc$ denotes the complex conjugate of the preceding terms. Substituting Eqs.~(\ref{eq:1 order solution}) into Eq.~(\ref{eq:2 order equation x}), we obtain
\begin{subequations}\label{eq:2 order equation No.2}
\begin{align}
\uD_0^2 x_2+(1-\kappa)x_2 &=-2i(1-\frac12\kappa) \uD_1A e^{i(1-\frac12\kappa)T_0},\label{eq:2 order equation No.2 x}\\
\uD_0^2 y_2+y_2 &=-2i\uD_1B e^{i T_0}.\label{eq:2 order equation No.2 y}
\end{align}
\end{subequations}
To eliminate secular terms of $x_2$ and $y_2$, we have
\begin{align}\label{eq:D_1AB}
\uD_1A=0\quad \&\quad \uD_1B=0.
\end{align}
Because we have included all the information of general solution in the first order solution Eqs.~(\ref{eq:1 order solution}), we should not consider general solution in higher order solutions, or else the coefficients will be underdetermined. Nayfeh in his book\cite{Nayfeh1995} has a further discussion. Hence, the general solution of the second order equations is 0. In addition, from Eqs.~(\ref{eq:D_1AB}) we know that particular solution of Eqs.~(\ref{eq:2 order equation No.2}) equals 0. Taking general and particular solutions together, we have
\begin{align}\label{eq:2 order solution}
x_2=y_2=0.
\end{align}
Substituting Eqs.~(\ref{eq:1 order solution}), Eqs.~(\ref{eq:D_1AB}) and Eq.~(\ref{eq:2 order solution}) into Eq.~(\ref{eq:3 order equation x}), we obtain
\begin{multline}
\label{eq:to solve x3}
\uD_0^2x_3+(1-\kappa)x_3 =
\big(-2i(1-\frac12\kappa)\uD_2A+2(1-2\kappa)A^2\bar{A}\big)e^{i(1-\frac12\kappa)T_0}\\
+(1-\kappa)\Big[(1-\kappa)A^3e^{i(3-\frac32\kappa) T_0}+AB^2e^{i(3-\frac12\kappa)T_0}+\bar{A}B^2e^{i(1+\frac12\kappa)T_0}\Big]+cc.
\end{multline}
To eliminate secular terms of $x_3$, we have
\begin{align}
\label{eq:solve A}
-i(1-\frac12\kappa)\uD_2A+(1-2\kappa)A^2\bar{A}=0.
\end{align}
Notice that $A$ is independent of $T_1$. For convenience, we write $A$ in the polar form
\begin{align}
A(T_2)=\frac12 a(T_2)e^{i\theta(T_2)},\label{eq:A polar form}
\end{align}
where $a$ and $\theta$ are real functions of $T_2$. Substituting  Eq.~(\ref{eq:A polar form}) into Eq.~(\ref{eq:solve A}) and separating the result into real and imaginary parts, we obtain
\begin{subequations}
\begin{align}
\uD_2 a&=0,\\
\uD_2 \theta &= -\frac14 a_0^2 (1-\frac32\kappa).
\end{align}
\end{subequations}
It follows that $a$ is a constant and hence
\begin{subequations}
\begin{align}
a&=a_0,\\
\theta&=-\frac14 a_0^2 (1-\frac32\kappa)T_2 +\theta_0,
\end{align}
\end{subequations}
where $a_0$ and $\theta_0$ are real constants. Returning to Eq.~(\ref{eq:A polar form}), we find
\begin{equation}
A(t)=\frac12a_0e^{i(-\frac{1-\frac32\kappa}{4}\varepsilon^2a_0^2 t+\theta_0)},\label{eq:A}
\end{equation}
where we have used $T_2=\varepsilon^2 t$. Similarly, we can deduce
\begin{equation}
B(t)=\frac12b_0e^{i(-\frac14\varepsilon^2b_0^2t +\phi_0)},\label{eq:B}
\end{equation}
where $b_0$ and $\phi_0$ are real constants. Substituting for $A$ and $B$ from Eqs.~(\ref{eq:A}) and (\ref{eq:B}) into Eq.~(\ref{eq:to solve x3}) and setting the general solution as 0, we obtain
\begin{multline}
x_3=\frac{1-\kappa}{4} \Big[\frac{(1-\kappa) a_0^3}{(1-\kappa)-9{\sigma_x}^2}e^{i(3\sigma_x t+3\theta_0)}
+\frac{ a_0b_0^2}{1-\kappa-(\sigma_x+2\sigma_y)^2}e^{i((\sigma_x+2\sigma_y)t+\theta_0+2\phi_0)}\\
+\frac{a_0b_0^2}{1-\kappa-(2\sigma_y-\sigma_x)^2}e^{i((2\sigma_y-\sigma_x)t-\theta_0+2\phi_0)}\Big]+cc, \label{eq:full x3}
\end{multline}
where
\begin{align*}
\sigma_x&\equiv(1-\frac12\kappa)-\frac{1-\frac32\kappa}{4}\varepsilon^2a_0^2
\simeq 1-\frac14\varepsilon^2a_0^2-\frac12\kappa,\\
\sigma_y&\equiv1-\frac14 \varepsilon^2 b_0^2.
\end{align*}
The parameters $\sigma_x$ and $\sigma_y$, as we will see in Eqs.~(\ref{eq:x & y final}), are just the primary peak frequencies in the spectra. Normally, $\sigma_x$ and $\sigma_y$  approximately equal 1, and $\kappa\ll 1$; thus the third term in Eq.~(\ref{eq:full x3}) is much greater than the others. We may neglect the small terms and simplify the solution as
\begin{align}\label{eq:short x3}
x_3= \frac14\frac{(1-\kappa) a_0b_0^2}{1-\kappa-(2\sigma_y-\sigma_x)^2}
e^{i((2\sigma_y-\sigma_x)t-\theta_0+2\phi_0)}
+cc.
\end{align}
Then
\begin{align}
x=&\varepsilon x_1 +\varepsilon^2 x_2 +\varepsilon^3 x_3 \notag\\
=&\frac12 a e^{i(\sigma_x t+\theta_0)}+\frac14  \frac{(1-\kappa)ab^2}{1-\kappa-(2\sigma_y-\sigma_x)^2}
e^{i((2\sigma_y-\sigma_x)t-\theta_0+2\phi_0)}
+cc,
\end{align}
where $a\equiv\varepsilon a_0$ and $b\equiv\varepsilon b_0$. Similarly, we can find solution for y. According to the initial condition $x'(0)=0,\,y'(0)=0$, we have $\theta_0=\phi_0=0$. Hence the solution is
\begin{subequations}
\label{eq:x & y final}
\begin{align}
x=a\cos\sigma_x t +\frac{(1-\kappa)ab^2}{2b^2-a^2 -4\kappa}\cos(2\sigma_y-\sigma_x)t \label{eq:x & y final x},\\
y=b\cos\sigma_y t +\frac{(1-\kappa)a^2b}{2a^2-b^2
+4\kappa}\cos(2\sigma_x-\sigma_y)t \label{eq:x & y final y}.
\end{align}
\end{subequations}
We find that $x$ (or $y$) oscillation is the superposition of two harmonic components with a small angular frequency difference $2(\sigma_y - \sigma_x)$. The superposition  of the two components results in an amplitude modulation whose period is given by
\begin{align}
T&=\frac{2\pi}{2\Delta\sigma}\simeq \frac{4\pi}{a^2-b^2+2\kappa},
\label{eq:T}
\end{align}
where $\Delta\sigma\equiv\sigma_y-\sigma_x$. When only consider the leading terms in Eqs.~(\ref{eq:x & y final x}) and (\ref{eq:x & y final y}), we may approximate the angular frequency difference between $x$ and $y$ oscillations as $\sigma_y - \sigma_x$. This means that the period of phase shift is $\frac{2\pi}{\Delta\sigma}=2T$, twice the period of the amplitude modulation as discussed above. And $\Delta\sigma$ can be seen as the average angular speed of procession.

\begin{figure}[t]
	\begin{center}
    \subfigure[]{\label{fig:comparison a}
    \includegraphics[width=0.2\textwidth]{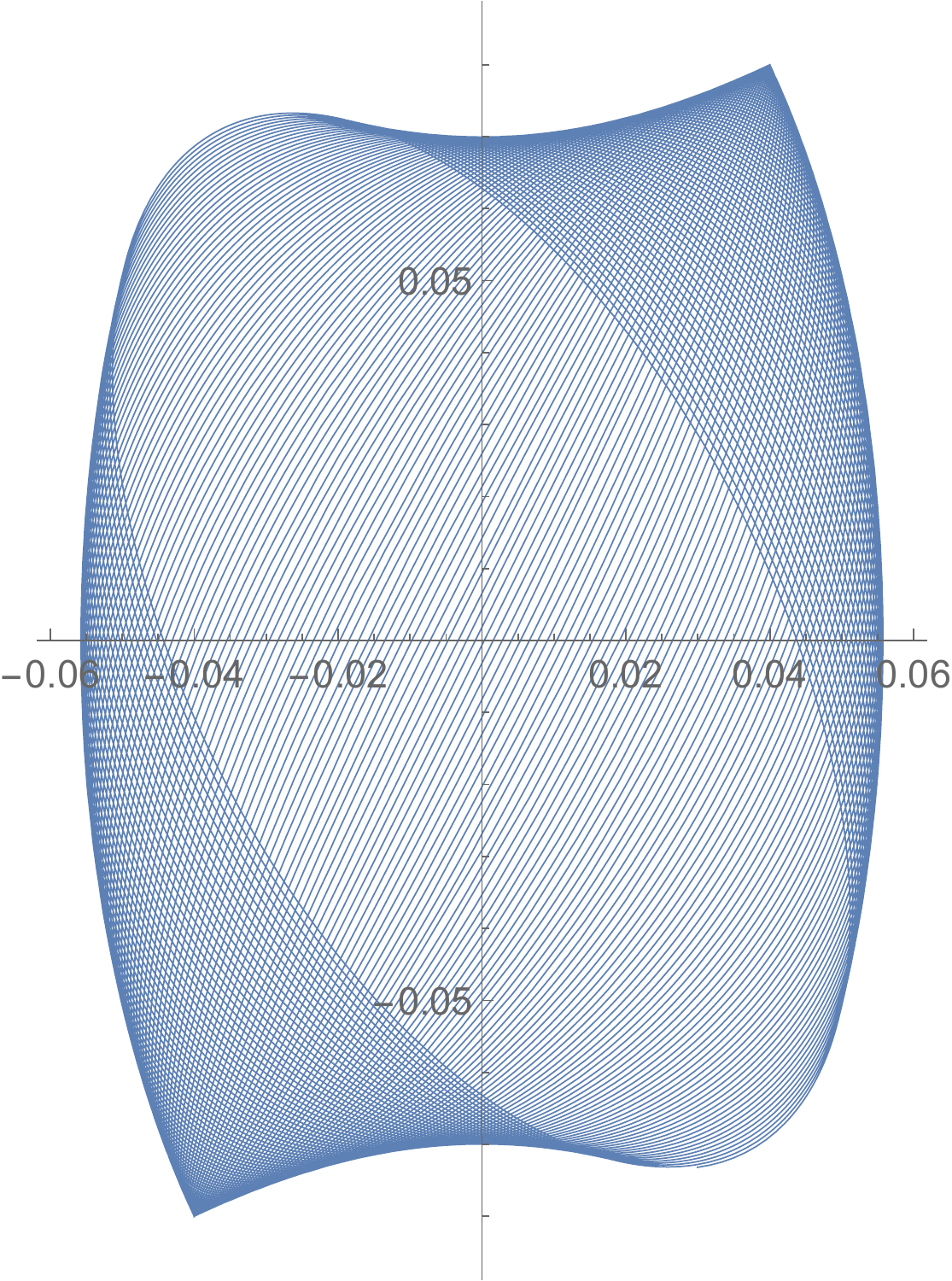}
    }
    \subfigure[]{\label{fig:comparison b}
    \includegraphics[width=0.2\textwidth]{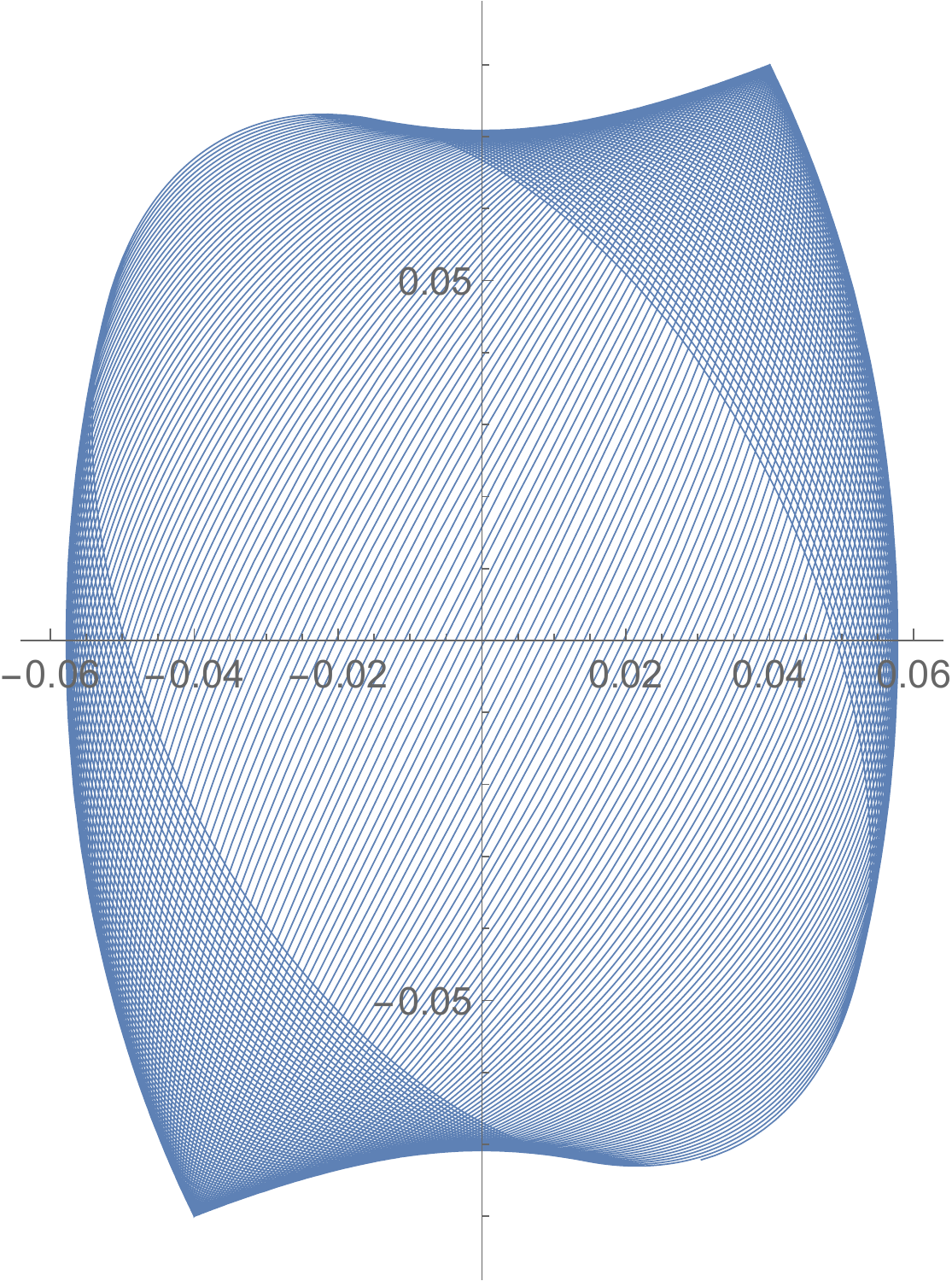}
    }
    \subfigure[]{\label{fig:comparison c}
    \includegraphics[width=0.4\textwidth]{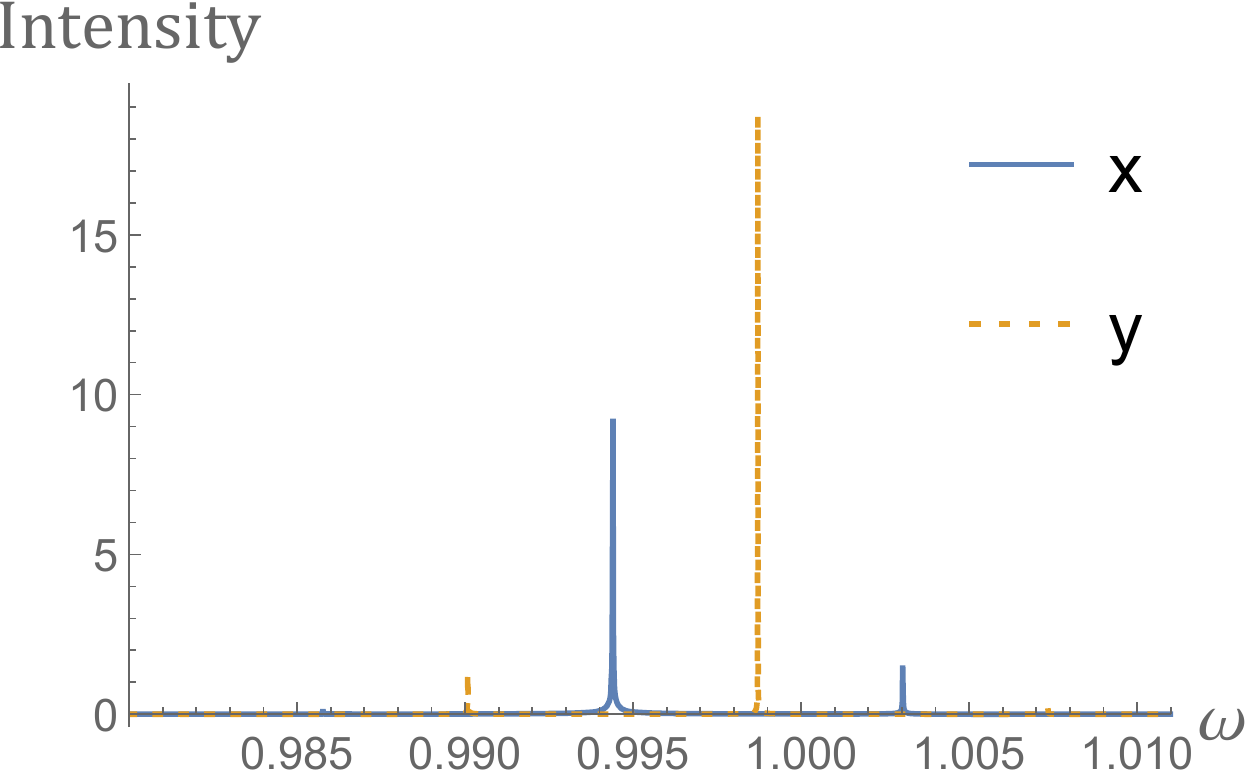}
    }

	\subfigure[]{\label{fig:comparison d}
    \includegraphics[width=0.2\textwidth]{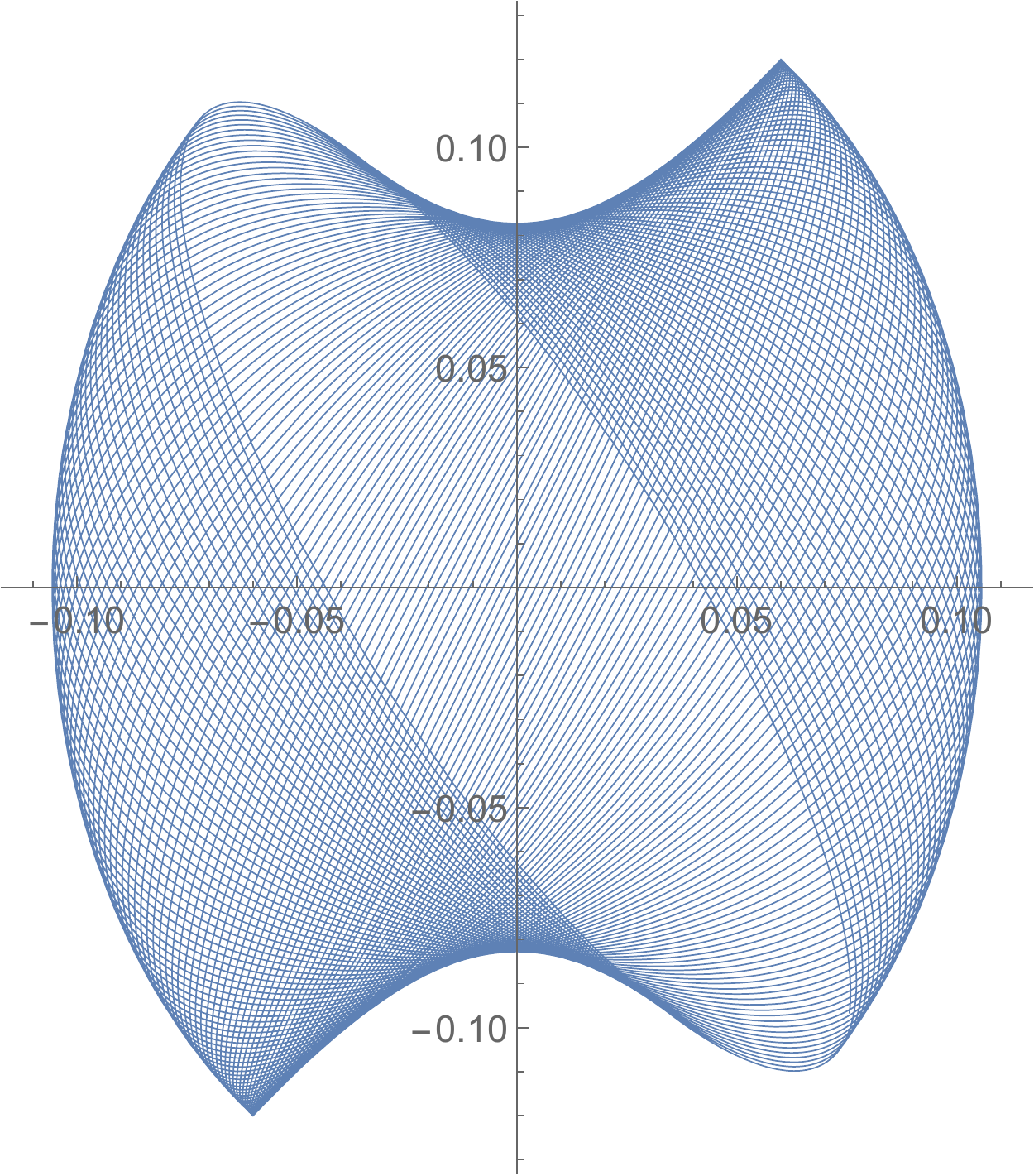}
    }
    \subfigure[]{\label{fig:comparison e}
    \includegraphics[width=0.23\textwidth]{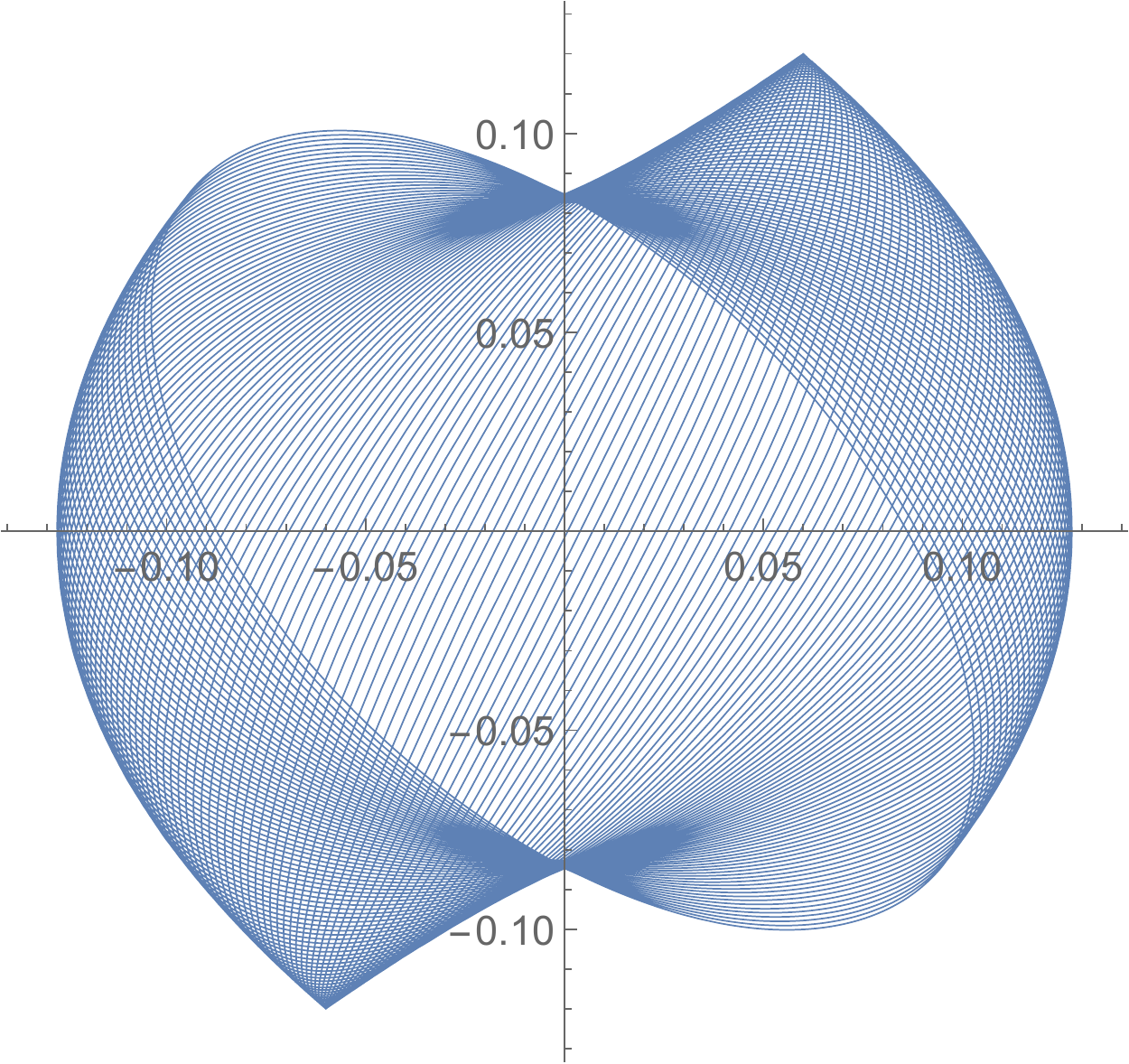}
    }
    \subfigure[]{\label{fig:comparison f}
    \includegraphics[width=0.4\textwidth]{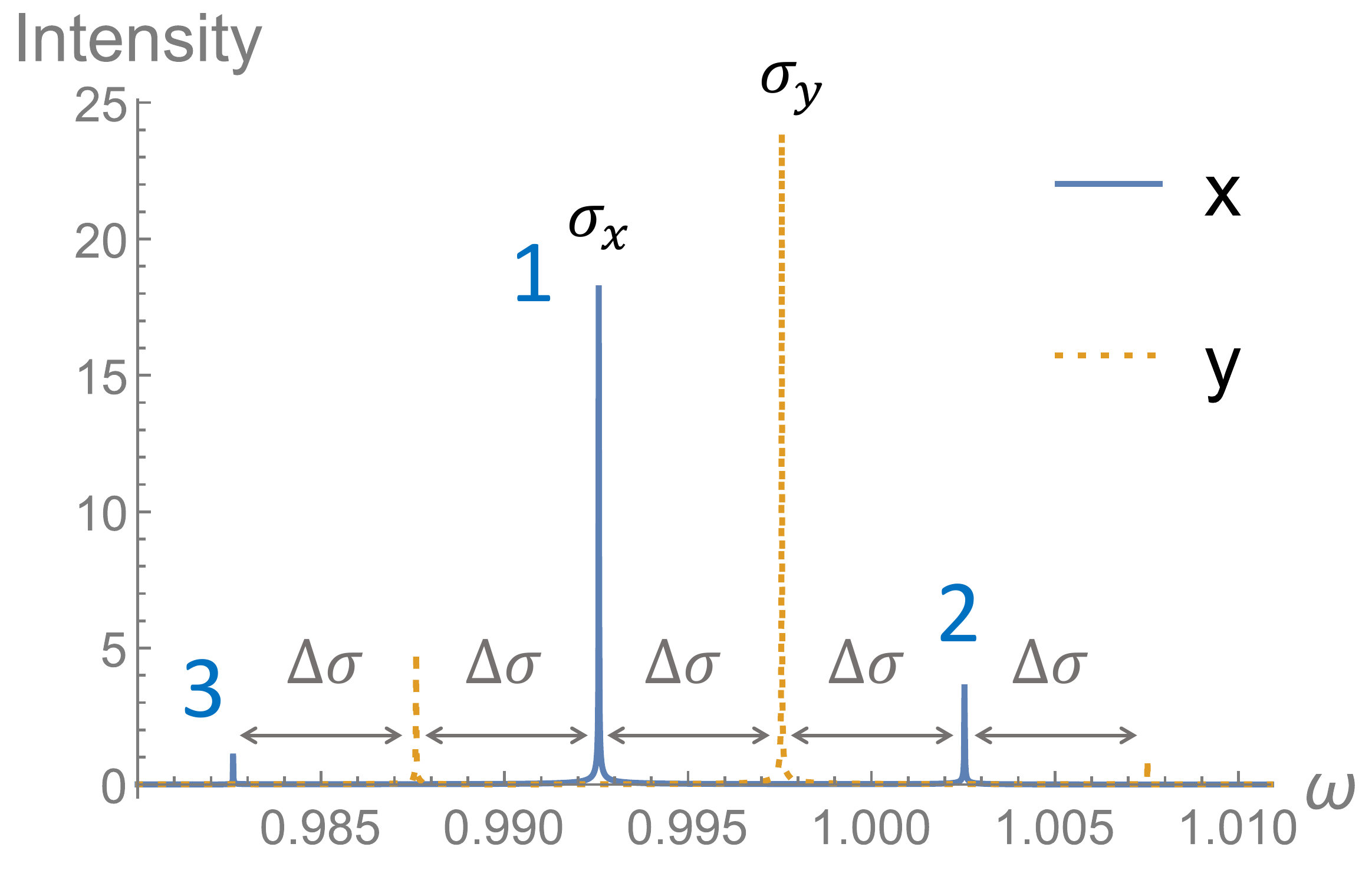}
    }
	\caption{(a) and (b) show the numerical and analytical trajectory diagrams for $\kappa=0.01,\,x_0=0.04,\,y_0=0.08$, respectively.  (c) is the frequency spectrum of (a). (d) and (e) show the numerical and analytical trajectory diagrams for $\kappa=0.01,\,x_0=0.06,\,y_0=0.12$, respectively. (f) is the frequency spectrum of (d). $\sigma_{x}$ and $\sigma_y$ denote the frequencies of the highest peaks of $x$ and $y$ motion.}
	\label{fig:comparison}
	\end{center}
\end{figure}

Notice that $a$ and $b$ are not the initial coordinates  $x_0$ and $y_0$. To acquire the initial position, we set $t=0$ in Eqs.~(\ref{eq:x & y final}) and obtain
\begin{subequations}
\begin{align}
x_0=a +\frac{(1-\kappa)a b^2}{2b^2-a^2 -4\kappa}\\
y_0=b +\frac{(1-\kappa)a^2 b}{2a^2-b^2
+4\kappa}
\end{align}
\label{eq:x0 & y0}
\end{subequations}
In analytical calculations, the parameters $(a, b)$ are solved from $(x_0,y_0) $ according to Eqs.~(\ref{eq:x0 & y0}). The solution is multi-valued and unrealistic solutions are eliminated. Thus, numerical and analytical results can be compared. We plot the simulative and analytical trajectories in Figs.~\ref{fig:comparison a} and \ref{fig:comparison b} for $\kappa=0.01,\,x_0=0.04,\,y_0=0.08$ in both numerical and analytical solutions. As we can see, there are only subtle difference between the two trajectories, see Figs.~\ref{fig:comparison a} and \ref{fig:comparison b}. However, when we take  $\kappa=0.01,$ and larger initial displacements $x_0=0.06,\,y_0=0.12$, the difference becomes obvious, see Figs.~\ref{fig:comparison d} and \ref{fig:comparison e}. Then we plot frequency spectra of the numerical results for these parameters, see Figs.~\ref{fig:comparison c} and \ref{fig:comparison f}. For each oscillation, there are several discrete peaks. The main and secondary peaks for each oscillation correspond to the two terms in analytical result Eqs.~(\ref{eq:x & y final x}) and (\ref{eq:x & y final y}). When the amplitude is small, as shown in Fig.~\ref{fig:comparison c}, only the secondary peak is competitive to the main peak in each oscillation, and the effect of other peaks are too weak to be observed. Under this circumstance, analytical results fit well with numerical results. However, when the amplitude is increased, as shown in Fig.~\ref{fig:comparison f}, the effect of other minor peaks are not negligible and the first order analytical solution will lose accuracy.

\begin{figure}[t]
	\begin{center}
    \subfigure[]{\label{fig:zetax_kappa}
	\includegraphics[width=0.4\textwidth]{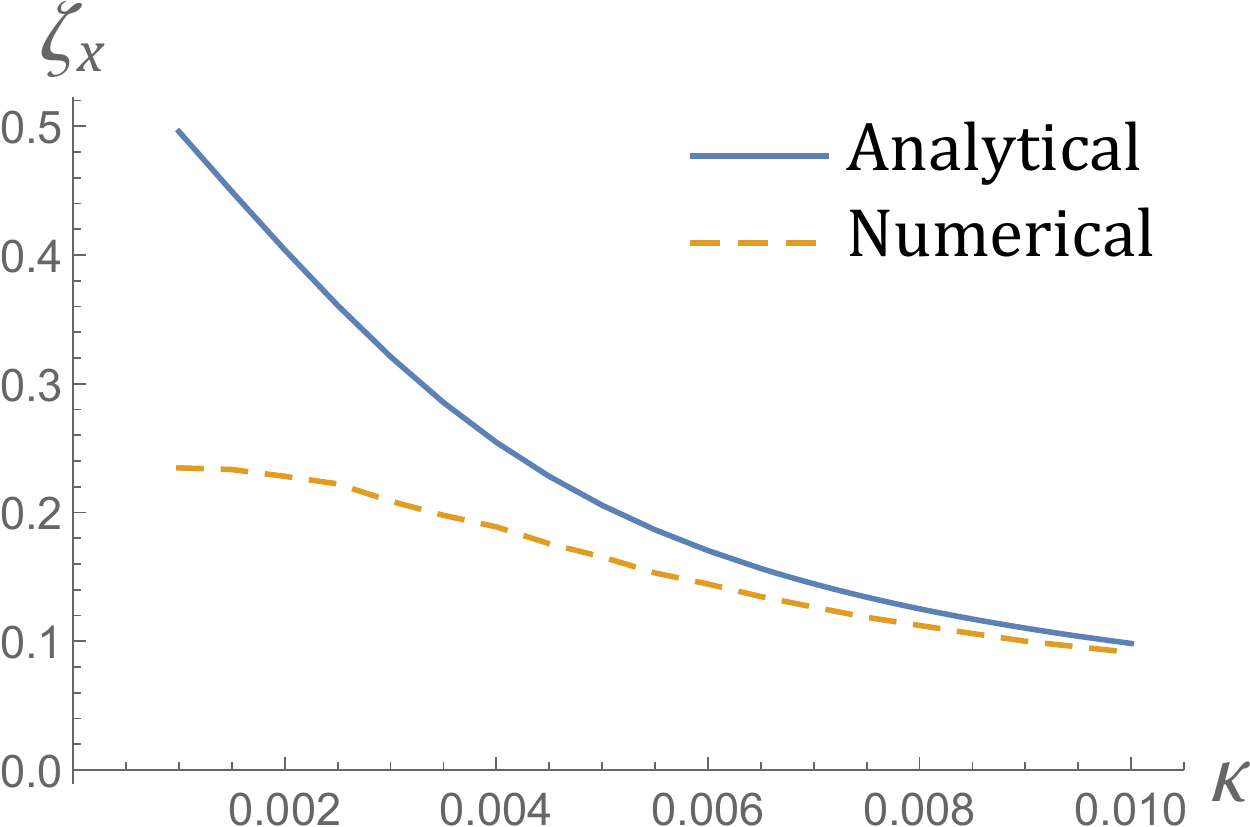}
    }
    \subfigure[]{\label{fig:sigmax1_kappa}
	\includegraphics[width=0.4\textwidth]{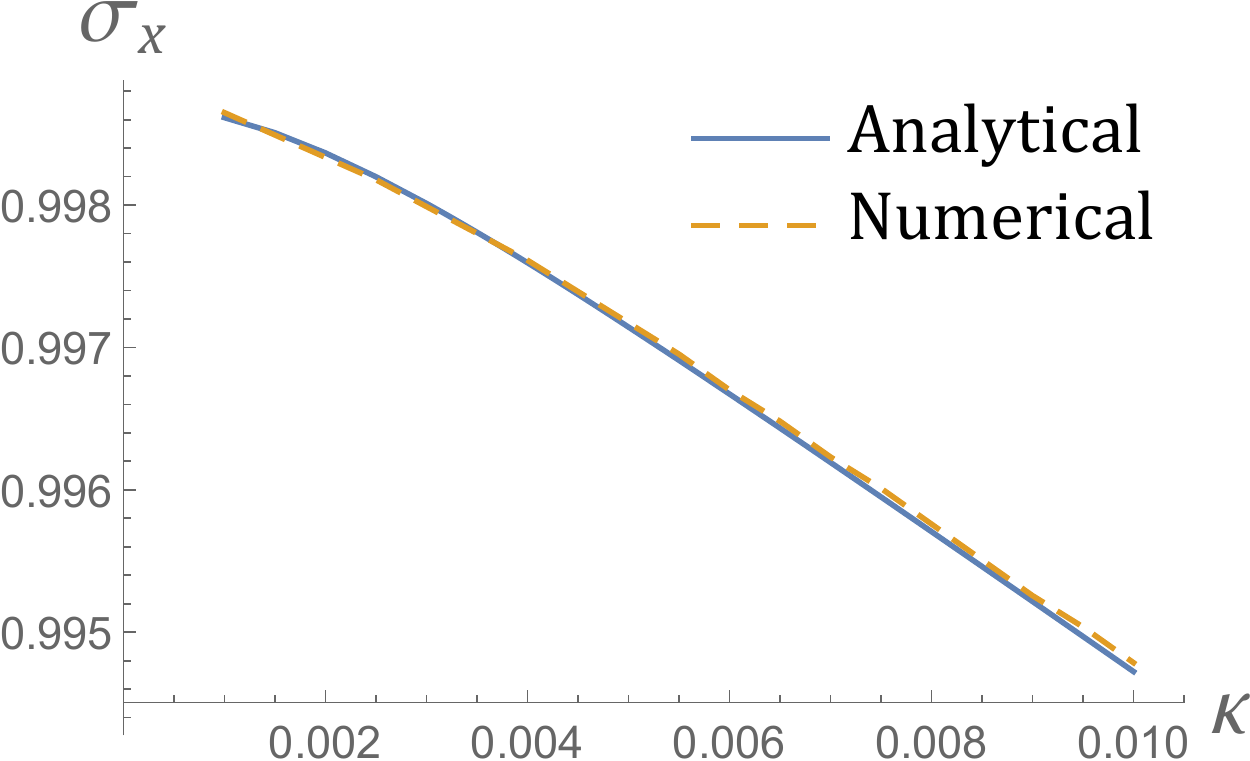}
    }
    \subfigure[]{\label{fig:sigmax2_kappa}
	\includegraphics[width=0.4\textwidth]{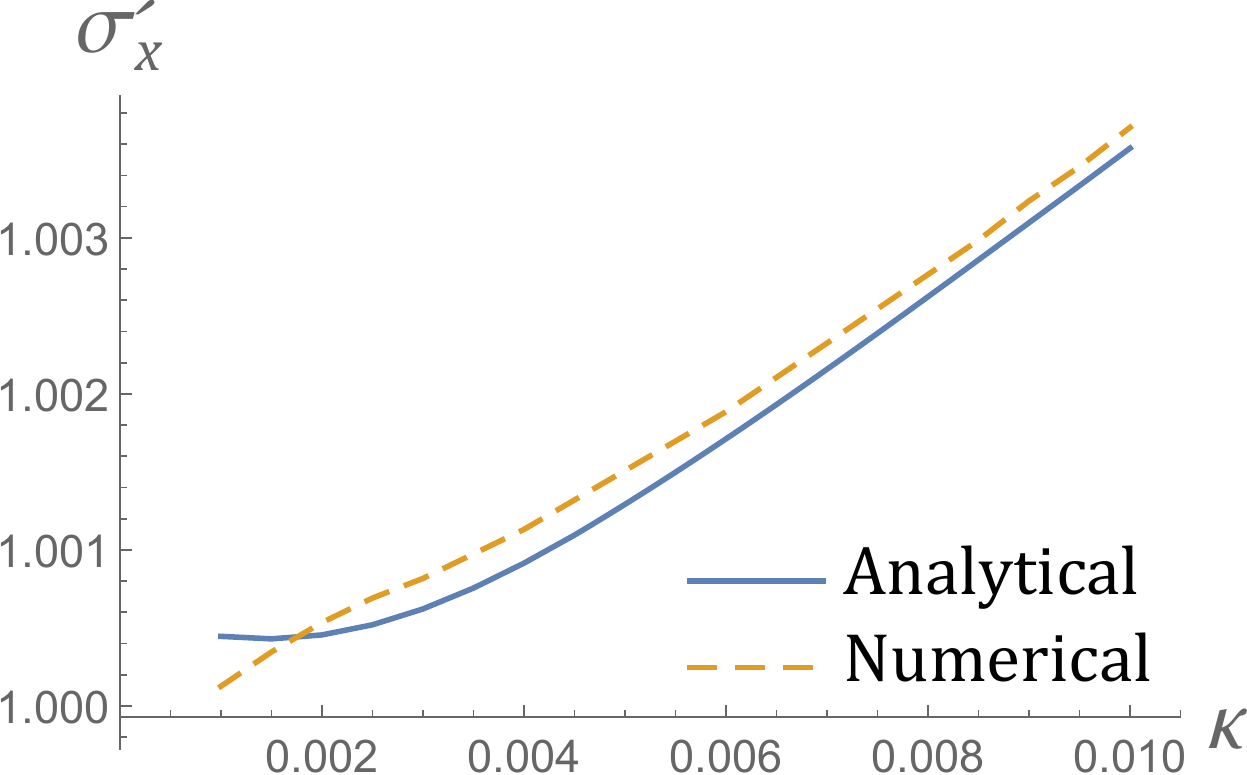}
    }
    \subfigure[]{\label{fig:deltasigma_kappa}
	\includegraphics[width=0.4\textwidth]{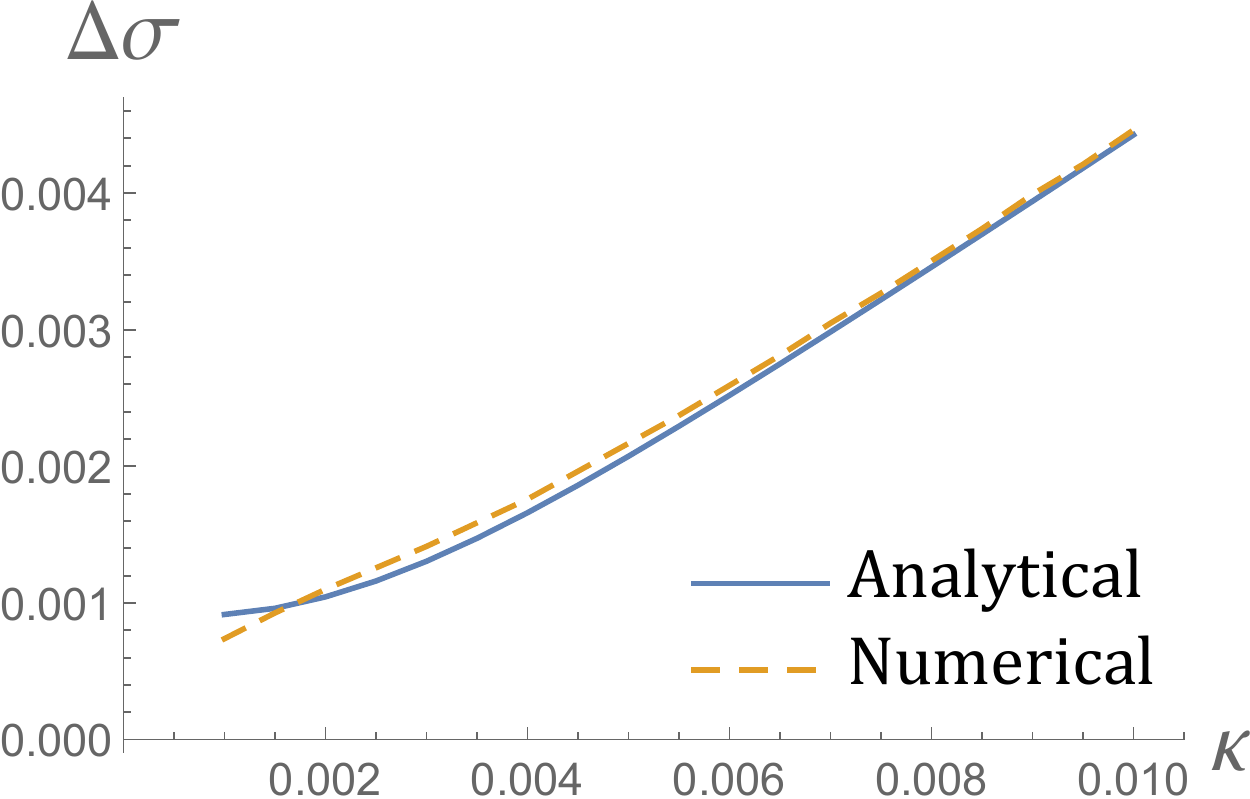}
    }
	\caption{Modulation depth $\zeta_x$, peak frequencies $\sigma_x$, $\sigma_x'$, and frequency difference $\Delta\sigma$ with respect to $\kappa$ for $x_0=0.03,\,y_0=0.06$.}
	\end{center}
\end{figure}
Now we study how coupling coefficient $\kappa$ and initial amplitudes affect the nonlinear coupling. To describe it more clearly, we label the intensities of primary and secondary peak as $h_x$ and $h_x'$ for $x$ oscillation. Also, we label the dimensionless frequencies of the primary and secondary peaks as $\sigma_x$ and $\sigma_x'$ ($\sigma_y$ and $\sigma_y'$) for $x$ ($y$) oscillation. Then, we introduce
\begin{align}
    \zeta_x \equiv h_x'/h_x.
\end{align}
If the motion only comprises two harmonic components, $\zeta_x$ is the modulation depth, i.e. the ratio of modulation amplitude to carrier amplitude in terms of amplitude modulation. In this case, the modulation depth $\zeta_x$ can be found from the ratio of the coefficients of the second term and first term in Eqs.~(\ref{eq:x & y final}):
\begin{align}
    \zeta_x=\frac{(1-\kappa)b^2}{2b^2-a^2 -4\kappa}.
\end{align}
Thus, we can verify the accuracy of the analytical results. Fig.~\ref{fig:zetax_kappa} shows how $\zeta_x$ depends on $\kappa$ for $x_0=0.03,\,y_0=0.06$, where $\kappa$ varies from 0.01 to 0.1. We find that  $\zeta_x$ becomes smaller as $\kappa$ increases, indicating that nonlinear effect is weaker for larger $\kappa$.

Moreover, we have studied how the main peak frequency $\sigma_x$ and secondary peak frequency $\sigma_x'$ change with $\kappa$. Observing Eqs.~(\ref{eq:x & y final}), we see that 
\begin{subequations}
\begin{align}
    \sigma_x'=2\sigma_y-\sigma_x =\sigma_x+2\Delta\sigma,\\
    \sigma_y'=2\sigma_x-\sigma_y=\sigma_y-2\Delta\sigma,
\end{align}
\end{subequations}
which is also verified by the numerical result, see Fig.~\ref{fig:comparison f}. To compare the analytical and numerical results, we have Figs.~\ref{fig:sigmax1_kappa}-\ref{fig:deltasigma_kappa}.  Although $\sigma_x$ and $\sigma_x'$ change very little with $\kappa$, the modulation frequency $\Delta\sigma$ increase drastically. This means that the average procession speed is significantly increases for greater $\kappa$.
\begin{figure}[t]
	\begin{center}
    \subfigure[]{\label{fig:zetax_alpha}
	\includegraphics[width=0.4\textwidth]{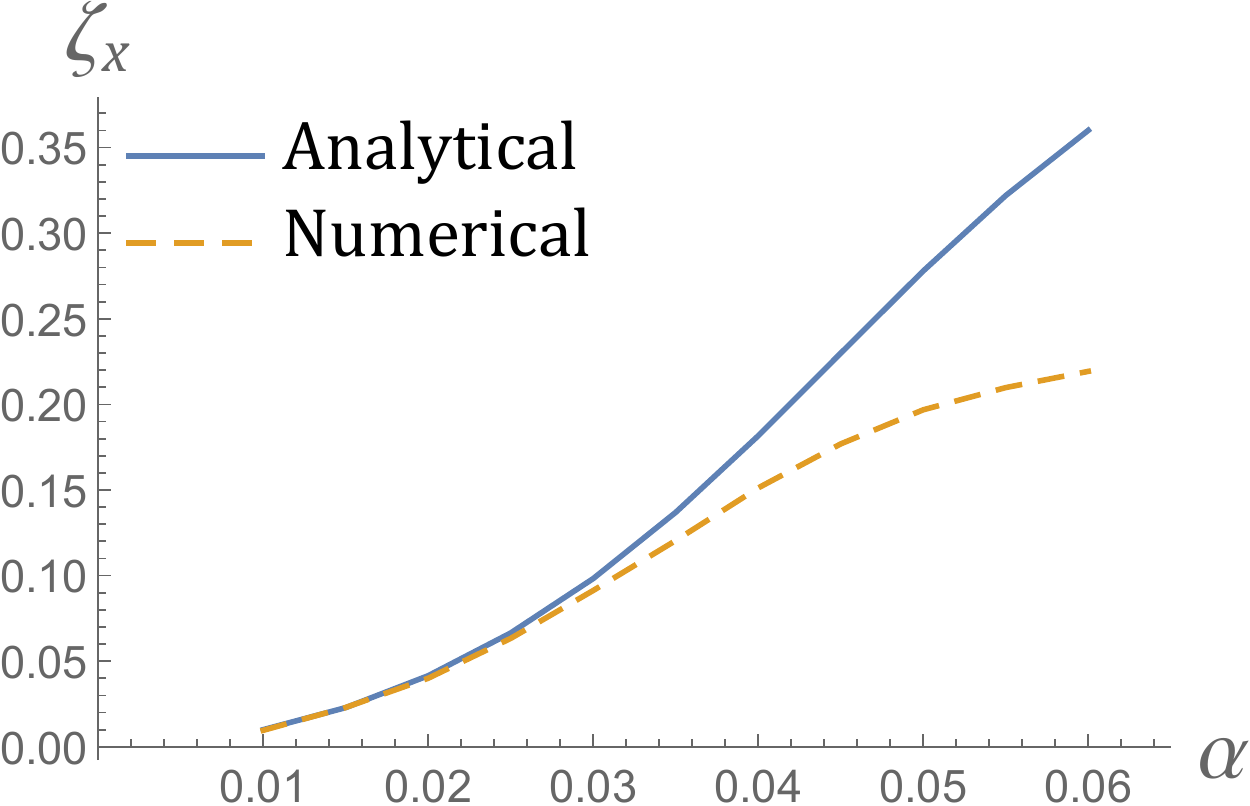}
    }
    \subfigure[]{\label{fig:sigmax1_alpha}
	\includegraphics[width=0.4\textwidth]{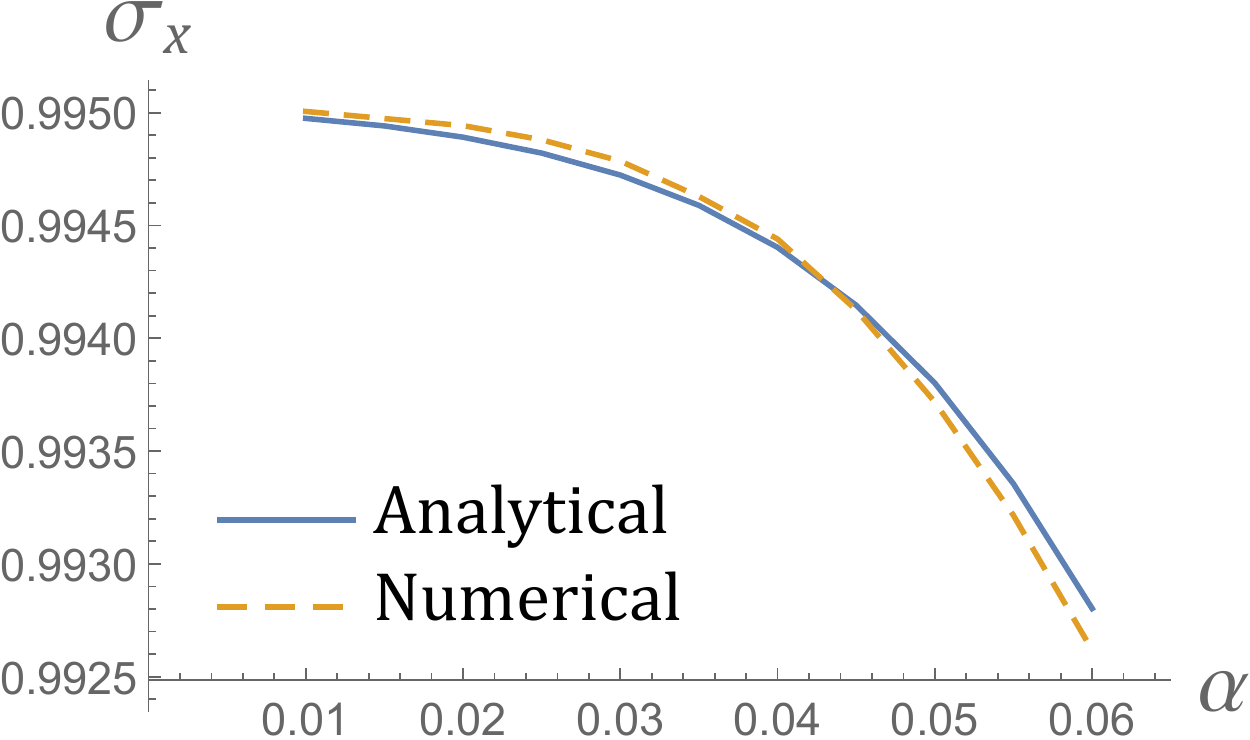}
    }
    \subfigure[]{\label{fig:sigmax2_alpha}
	\includegraphics[width=0.4\textwidth]{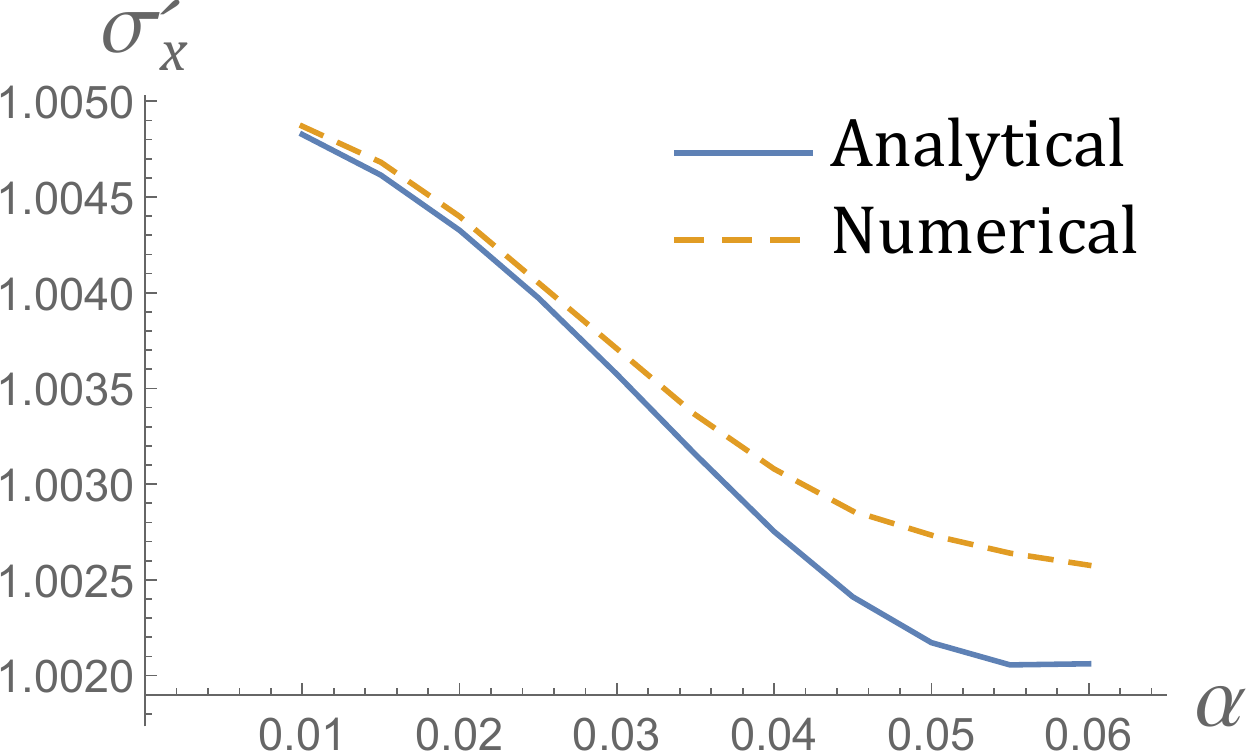}
    }
    \subfigure[]{\label{fig:deltasigma_alpha}
	\includegraphics[width=0.4\textwidth]{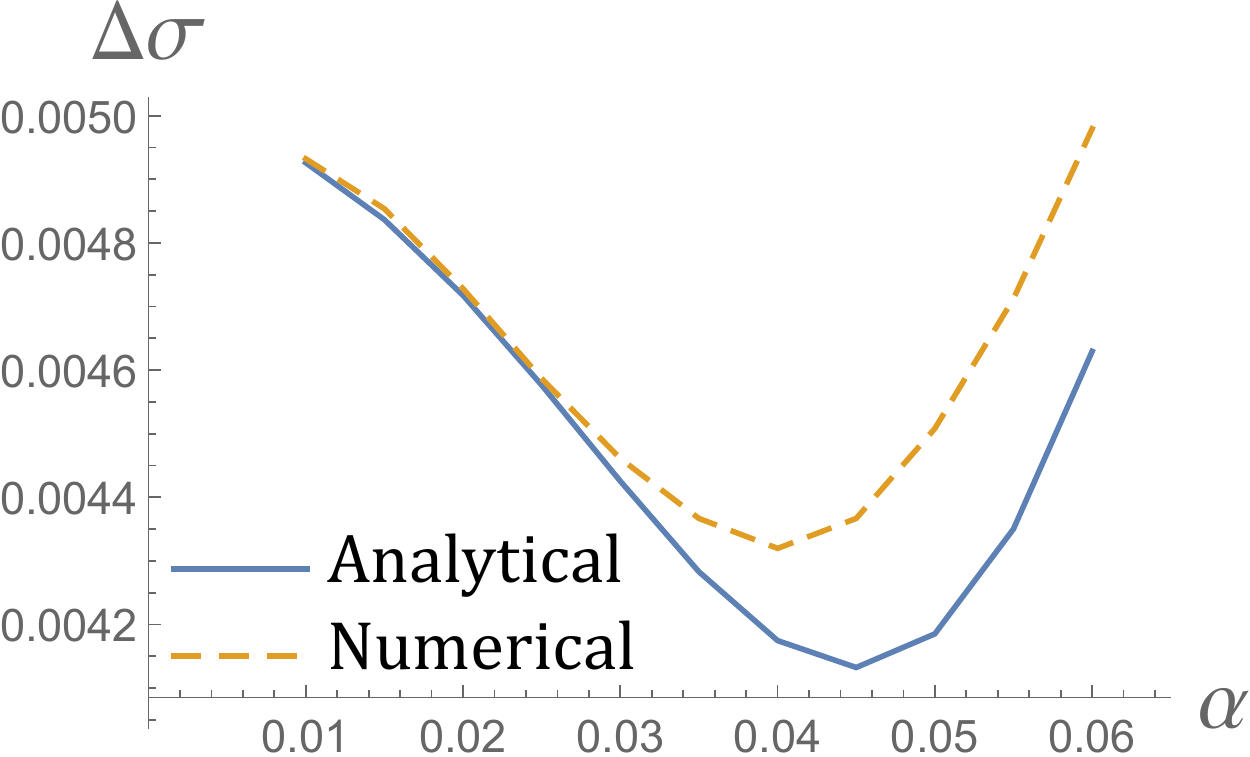}
    }
    \label{fig:change with alpha}
	\caption{Modulation depth $\zeta_x$, peak frequencies $\sigma_x$, $\sigma_x'$, and frequency difference $\Delta\sigma$ with respect to $\alpha$ for $\kappa=0.01,\,x_0=\alpha,\,y_0=2\alpha$.}
	\end{center}
\end{figure}

To see how amplitudes affect the pendulum motion, we plot modulation depth $\zeta_x$, peak frequencies $\sigma_x$ and $\sigma_x'$ as well as frequency difference $\Delta\sigma$ in Figs.~\ref{fig:zetax_alpha}-\ref{fig:deltasigma_alpha} as functions of initial positions. We take $\kappa=0.01$, and initial positions $x_0=\alpha,\,y_0=2\alpha$, and $\alpha$ varies from 0.01 to 0.06. In Figs.~\ref{fig:sigmax1_alpha} and \ref{fig:sigmax2_alpha}, both $\sigma_x$ and $\sigma_x'$ decrease as $\alpha$ increases, but the modulation frequency $\Delta\sigma$ is nonmonotonic: it reaches a minimum near $\alpha=0.04$ in Fig.~\ref{fig:deltasigma_alpha}. Fig.~\ref{fig:zetax_alpha} shows that the modulation depth is greater for larger $\alpha$, as nonlinearity is stronger for larger amplitudes. The analytical results also gradually lose accuracy for larger amplitudes due to stronger nonlinearity. The regular motion we have solved is valid only in the range of weak nonlinearity. In addition, when the rod's natural frequency is comparable to pendulum's natural frequency, the pendulum’s motion will become far more complicated than we have presented. Finally, we have carried an experiment to verify the dependence of modulation frequency on the coupling coefficient and amplitudes. The experiment data included in supplementary material 2\cite{supplementary_material_2} match our theoretical trends well. 

\section{Conclusion}
The phenomenon of conversion between radial and azimuthal oscillations described in this paper is common for asymmetric pendulums. However, nonlinear coupling between the two oscillations is usually overlooked. We explain it as amplitude modulation due to nonlinear coupling. The pendulum’s motion patterns are solved numerically and analytically. The amplitude modulation period $T$ is explicitly expressed in terms of coupling coefficient and amplitudes. The amplitude dependence of $T$ is a typical nonlinear behavior. The advantage of this experimental apparatus is that it has appealing visual effects and the strength of coupling and other oscillation parameters are controllable thus can easily be compared to theoretical results. The method of multiple scales we introduce can easily be followed by undergraduate students. This work provides a good demonstration as well as a research project of nonlinear dynamics on different levels.

\section{Acknowledgment}
The authors are grateful to Mr. Lintao Xiao for instructive discussion and proofreading.



\begin{thebibliography}{99}

\bibitem{Whitaker2004}
Whitaker and J.~Robert, ``Types of Two-Dimensional Pendulums and Their Uses in Education," \href{https://doi.org/10.1023/B:SCED.0000041830.98845.5f}{Science \& Education} \textbf{13}, 401-415 (2004).

\bibitem{Greenslade2003}
T. B. Greenslade, ``Devices to Illustrate Lissajous Figures," \href{http://dx.doi.org/10.1119/1.1607806}{Phys. Teach.} \textbf{41(41)}, 351-354 (2003).

\bibitem{Whitaker1991}
R. J. Whitaker, ``A note on the Blackburn pendulum," \href{https://doi.org/10.1119/1.16543}{Am. J. Phys.} \textbf{59(4)}, 330-333 (1991).

\bibitem{Jorge2011}
Jorge Quereda \textit{et al.}, ``Calibrating the frequency of tuning forks by means of Lissajous figures," \href{https://doi.org/10.1119/1.3546095}{Am. J. Phys.} \textbf{79(5)}, 517 (2011).

\bibitem{Singh2018}
P. Singh \textit{et al.}, ``Study of normal modes and symmetry breaking in a two-dimensional pendulum,''  eprint: \href{https://arxiv.org/abs/1806.06222}{arXiv:1806.06222} (2018).

\bibitem{Feynman}
Richard P. Feynman, \textit{The Feynman lectures on physics} (Reading, Mass. :Addison-Wesley Pub. Co., Boston, 1918-1988).

\bibitem{supplementary_material_1}
Supplementary material 1: an experiment video clip and an animation of the asymmetric pendulum. $<$\url{https://box.nju.edu.cn/d/5777b0ed741e43a192f9/}$>$.

\bibitem{IYPT}
IYPT 2018 Problem 11: Azimuthal-radial pendulum. $<$\url{https://www.iypt.org/problems/problems-for-the-31st-iypt-2018}$>$.

\bibitem{carrera2011}
Erasmo Carrera, Gaetano Giunta and Marco Petrolo, \textit{Beam Structures: Classical and Advanced Theories} (Wiley, Chichester, 2011).

\bibitem{NDSolve}
Wolfram Mathematica document about NDSolve method. $<$\url{https://reference.wolfram.com/language/ref/NDSolve.html}$>$.
  
\bibitem{Sturrock}
P. A. Sturrock and G. P. Thomson, ``Non-linear effects in electron plasmas,'' \href{https://doi.org/10.1098/rspa.1957.0176}{R. Soc. Lond. A} \textbf{242}, 277–299 (1957).

\bibitem{Nayfeh1993}
A. H. Nayfeh, \textit{Introduction to Perturbation Techniques} (Wiley-VCH, Zurich, 1993).

\bibitem{Nayfeh1995}
A. H. Nayfeh and D. T. Mook, \textit{Nonlinear Oscillations}, Wiley Classics Library Edition. (Wiley, Hoboken, 1995).

\bibitem{Nayfeh2000}
A. H. Nayfeh, \textit{Perturbation Methods} (Wiley-VCH, Zurich, 2000).

\bibitem{Kevorkian1996}
J.K. Kevorkian, J.D. Cole, \textit{Multiple Scale and Singular Perturbation Methods} (Springer, New York, 1996).

\bibitem{Bender1999}
Carl M. Bender, Steven A. Orszag, \textit{Advanced Mathematical Methods for Scientists and Engineers I} (Springer, New York, 1999).

\bibitem{supplementary_material_2}
Supplementary material 2: experiment data to verify the trend of the analytical results. $<$\url{https://box.nju.edu.cn/d/a497b5bc79be47d28e2a/}$>$.

\end{thebibliography}
\end{document}